\documentclass[twocolumn]{aastex631}
\twocolumngrid

\usepackage{subfigure}
\usepackage{wrapfig}
\usepackage{xcolor}

\begin{document}

\title{A New Measure of Assembly Bias using the Environment Dependence of the Luminosity Function}
\shorttitle{A New Measure of Assembly Bias}

\author[0009-0002-6750-4919]{Yikun Wang}
\affiliation{Department of Physics, Case Western Reserve University, 10900 Euclid Avenue, Cleveland, OH 44106-1715, USA}

\author[0000-0001-8286-6024]{Idit Zehavi}
\affiliation{Department of Physics, Case Western Reserve University, 10900 Euclid Avenue, Cleveland, OH 44106-1715, USA}

\author[0000-0001-7511-7025]{Sergio Contreras}
\affiliation{Donostia International Physics Center, Manuel Lardizabal Ibilbidea, 4, 20018 Donostia, Gipuzkoa,  Spain}

\author[0000-0002-5954-7903]{Shaun Cole}
\affiliation{Institute for Computational Cosmology, Department of Physics,
  Durham University, South Road, Durham, DH1 3LE, UK}

\author[0000-0002-5875-0440]{Peder Norberg}
\affiliation{Institute for Computational Cosmology, Department of Physics,
  Durham University, South Road, Durham, DH1 3LE, UK}
\affiliation{Centre for Extragalactic Astronomy, Department of Physics,
  Durham University, South Road, Durham, DH1 3LE, UK}

\correspondingauthor{Idit Zehavi}
\email{idit.zehavi@case.edu}

\begin{abstract} 

Assembly bias is the variation in the clustering of dark matter halos and galaxies that arises from correlations between the halo assembly history and the large-scale environment at fixed halo mass. In this work, we use the cosmological magneto-hydrodynamical simulation TNG300 to investigate how assembly bias affects the environment-dependent galaxy luminosity function. We measure the luminosity functions in bins of large-scale environment for the original simulated galaxy sample and for a shuffled sample, where the galaxies are randomly reassigned among halos of similar mass to remove assembly bias. By comparing them, we find distinct signatures, showing variations in the number of galaxies at the $\sim10\%$ level across all luminosities. Assembly bias increases the tendency of galaxies to reside in denser environments and further dilutes underdense regions, beyond the trends governed by halo mass. When separating by color, we see that assembly bias has a much bigger effect on red galaxies fainter than $\mathrm{M_r} - 5 \log{h} = -18.5$, which accounts for a $\sim20\%$ increase in the number of galaxies in the densest environment and a remarkable $50\%$ decrease in the least dense regions. The ratio of these measurements for the densest and least dense regions provides a significant assembly bias signal for the faint red galaxies, larger than a factor of two. Overall, our results provide a novel sensitive  measure of assembly bias, offering valuable insight for modeling the effect and a potential new route to detect it in observations.

\end{abstract}

\keywords{Dark matter --- Galaxies --- Galaxy dark matter halos --- Galaxy evolution --- Galaxy Formation --- Galaxy luminosities --- Hydrodynamical simulations --- Large-scale structure of the universe --- Luminosity function --- Observational cosmology}

\section{Introduction}
\label{sec:intro}

In the standard cosmological framework of hierarchical structure formation, galaxies reside inside dark matter halos and evolve alongside them (e.g., \citealt{1996clss.conf..349W, 2018ARA&A..56..435W}). The formation and evolution of the dark matter halos are primarily governed by gravitational interactions, allowing for accurate predictions through high-resolution numerical simulations. The formation of galaxies, however, is more complex, and depends on the detailed physical processes. Investigating the relation between galaxies and halos is essential for advancing our understanding of galaxy formation and for constraining cosmological models. Having a detailed understanding of the galaxy-halo connection is crucial to optimally use galaxies as a cosmological probe in order to take full advantage of the pristine measurements made by the next generation of galaxy surveys.

Important questions in this context are how does environment impact the connection between galaxies and halos and how are galaxies affected by the assembly history of the halos. A central assumption that provided the basis for the development of halo modeling is that the galaxy content of halos depends only on the halo mass and is statistically independent of the large-scale environment. This assumption originates from the uncorrelated nature of random walks describing halo assembly in the standard excursion set formalism \citep{1991ApJ...379..440B, 1999Ap&SS.267..355W, 1999MNRAS.302..111L}. In this picture, the variation of galaxy properties with large-scale environment (for example, the fraction of blue and red galaxies), is fully derived from the change of the halo mass function in these environments.

This assumption has been challenged, however, by the demonstration in numerical simulations that the clustering of halos of fixed mass depends on secondary halo properties such as age, concentration, spin and tidal anisotropy (e.g., \citealt{10.1111/j.1365-2966.2004.07733.x, 2005MNRAS.363L..66G, 2006ApJ...652...71W, 10.1111/j.1745-3933.2007.00292.x, 10.1093/mnras/sty109, 2019MNRAS.489.2977R, 2019MNRAS.487.1570S, 2025A&A...695A.159M}), a multi-faceted effect commonly referred to as {\it halo assembly bias}. If galaxy properties correlate with halo formation history, we expect the galaxy content of halos to also depend on the secondary halo properties and on large-scale environment, leading to distinct variations in the halo occupation functions \citep{2006ApJ...639L...5Z,2018ApJ...853...84Z, 2018MNRAS.480.3978A, 2019MNRAS.490.5693B}. The predictions for these so-called {\it occupancy variations} have shown that halos in denser environments more readily host galaxies at lower halo mass. 

The combined impact of halo assembly bias and occupancy variations lead to a change of the galaxy correlation function amplitude on large scales \citep{2007MNRAS.374.1303C}. This imprint of assembly bias on the galaxy distribution is often referred to as {\it galaxy assembly bias} (GAB hereafter). It is commonly studied in simulations by comparing the galaxy correlation function to that of a shuffled galaxy sample, where galaxies are randomly reassigned to halos of the same mass, effectively removing assembly bias (e.g., \citealt{2007MNRAS.374.1303C, 2008ApJ...686...41Z, 2016MNRAS.460.3100C, 2018ApJ...853...84Z, 2019ApJ...887...17Z, 2019MNRAS.484.1133C, 2021MNRAS.504.5205C, 2021MNRAS.507.4879X, 2021MNRAS.502.3242X,2020MNRAS.493.5506H,2023MNRAS.524.2507H}). Neglecting these effects can have important implications for the interpretation of galaxy clustering (e.g., \citealt{2014MNRAS.443.3044Z}). 

The observational evidence for GAB, however, remains inconclusive and controversial, given the challenge of constraining it without direct measures of halo properties. Several suggestive detections have been put forth (e.g., \citealt{2013MNRAS.433..515W, 2015MNRAS.452.1958H, 2020JCAP...10..058O, 2021MNRAS.502.3582Y, 2023MNRAS.525.3149C, 2024arXiv241111830O, 2024ApJ...974...29O, 2024ApJ...963..116P}),  while other studies indicate the impact of assembly bias to be small \citep[e.g.,][]{2007ApJ...664..791B, 2016ApJ...819..119L, 2017MNRAS.470..551Z, 2019MNRAS.488..470W, Kakos2024} and that previous claims were plagued by systematics \citep{2015MNRAS.452..444C, 2016MNRAS.457.4360Z, 2017MNRAS.471.1192S, 2019MNRAS.490.4945S}.

The contradictory results and lack of consensus motivate exploring alternative
measures of assembly bias. While previous studies have typically explored GAB in the context of galaxy clustering, here we set to explore how assembly bias manifests in the environment dependence of the galaxy luminosity function. The galaxy luminosity function (LF) measures the number density of galaxies as a function of luminosity and is frequently obtained for redshift surveys (e.g., \citealt{1988MNRAS.232..431E, 1996JRASC..90..337L, 2002MNRAS.336..907N, 2003ApJ...592..819B, 2012MNRAS.420.1239L}). It is a fundamental statistic that quantifies the populations of galaxies in the observable Universe, reflecting the impact of different physical processes that affect galaxies (e.g., \citealt{1991ApJ...379...52W, 2003ApJ...599...38B}). Computing how the LF varies with environment can help identify the environmental processes involved in galaxy formation and evolution.

It has been well established that the local environment can have a substantial role in shaping galaxy properties (e.g., \citealt{1974ApJ...194....1O, 1980ApJ...236..351D, 2004MNRAS.353..713K, 2005ApJ...629..143B}). The LF serves as a useful way to characterize galaxy populations and their environmental dependence. The advent of large galaxy surveys has enabled to measure it reliably over a range of environments. Early studies investigated how the LF varies between cluster, group and field regions or for voids \citep{10.1046/j.1365-8711.2003.06510.x, Hoyle_2005, 2006ApJ...652.1077R, 2008MNRAS.386.2285C}.

\citet{2015MNRAS.448.3665E} extended these analyses to different cosmic web environments, finding distinct variations attributed to the local-density dependence.  Other studies explored the LF as a function of galaxy density spanning the full range of environments using various observational data. It has been measured for galaxy surveys such as 2dF, GAMA and SDSS \citep{2005MNRAS.356.1155C,10.1093/mnras/stu1886, 2018MNRAS.476..741D}, and most recently also with early DESI data (S.\ Moore et al., in prep.). These studies have all found a consistent environment dependence, with the LF varying smoothly with density. Splitting the samples further by color or galaxy type leads to significant differences as well. However, the question remains whether these variations fully arise from the expected dependence on halo mass (i.e.\ the change of the halo mass function with environment) or if there is additional impact from the assembly history of the halos or the environment at large.   

\citet{10.1093/mnras/stu1886} compare the GAMA measurements to predictions of this statistic in a semi-analytic model, finding qualitative agreement but some notable discrepancies attributed to the galaxy formation model used. \citet{2018MNRAS.476..741D} measure the environment dependence of the LF in the SDSS, finding broad agreement with predictions of an abundance matching model based on the maximum circular velocity. We stress that none of these studies examined specifically or attempted to infer the assembly bias impact.

\citet{2004MNRAS.349..205M} use a halo occupation model, based on the conditional luminosity function model of \citet{2003MNRAS.339.1057Y}, to predict the dependence of the galaxy luminosity function on environment. This model assumes that the properties of a galaxy depend only on its dark matter halo mass, such that the change of the LF with large-scale environment is entirely due to the change of the halo mass function with environment. This can potentially allow one to look for secondary environmental effects by comparing the model predictions to observations. However, it is non-trivial in practice as deviations may be hard to discern or interpret, and assembly bias effects may enhance the same trends rather then produce distinct signatures. 

In this paper, we explore the environmental effects beyond halo mass and the signature of assembly bias in the environment dependence of the luminosity function, using the Illustris-TNG300 magneto-hydrodynamical simulation  \citep{2018MNRAS.475..676S}. In contrast to previous work, we aim to probe the impact of assembly bias directly, and assess the possibility of utilizing this statistic as a new way to detect it.  Using the galaxy and halo samples from the simulation, we first compute the galaxy LF in bins of large-scale environment. To infer the impact of assembly bias, we compare to the analogous measure for {\it shuffled} galaxy samples, where assembly bias is explicitly removed. We find significant signatures of assembly bias, especially in the extreme environment bins.  We also study color-selected galaxy samples and find particularly strong effects for faint red galaxies in the least dense regions.  These can serve as a guide to future modeling efforts and attempts to detect it in the real Universe. To our knowledge, this is the first direct prediction of the impact of assembly bias on the galaxy luminosity function.

The outline of the paper is as follows. In Section~\ref{sec:methods}, we describe the simulation and the methods used for measuring the environment dependence of the luminosity function and the effects of assembly bias. The main results of our analysis for the full galaxy sample are presented in Section~\ref{sec:result1}, while the results for color-selected samples are shown in Section~\ref{sec:result2}. We summarize our findings in Section~\ref{sec:concl}.  Appendix~\ref{sec:contam} discusses the impact of using the environment of the shuffled samples, while Appendices~\ref{sec:HMF} and~\ref{sec:slice} provide a few supplementary plots.

\section{Methods}
\label{sec:methods}

\subsection{The Illustris-TNG300 Simulation}
\label{subsec:TNG}

In this work, we utilize galaxy and halo samples from the Illustris-TNG300 simulation (hereafter TNG300). The simulation is part of ``The Next Generation'' Illustris suite of hydrodynamical cosmological simulations of galaxy formation \citep{10.1093/mnras/stx3112, 2018MNRAS.475..676S, 2018MNRAS.480.5113M, 2019ComAC...6....2N}, an improvement upon the original Illustris simulation (\citealt{2014MNRAS.444.1518V}, \citealt{2014MNRAS.445..175G}, \citealt{2015MNRAS.452..575S}). The suite includes three volumes and $18$ simulations in total, each of which has different physical sizes, mass resolutions or physical models\footnote{\url{https://www.tng-project.org/}}. 

We use the largest physical simulation box implementing the fiducial TNG model, TNG300, which has a side length of $L = 205 h^{ -1}\text{Mpc}$, with periodic boundary conditions. Specifically, we use TNG300-1, the main high-resolution run including the full physics model. This is one of the largest high-resolution hydrodynamical simulation publicly available to date. The simulation tracks the evolution of $2500^3$ gas cells and $2500^3$ dark matter particles from redshift $z = 127$ to $z = 0$, with a baryon mass resolution of $7.6 \times 10^6 h^{-1}M_{\odot} $ and a dark matter particle mass of $4.0 \times 10^7 h^{-1}M_{\odot} $. The cosmological parameters assumed in the simulation, $\Omega_M = 0.3089$, $\Omega_b = 0.0486$, $\sigma_8 = 0.8159$, $n_s = 0.9667$, and $h = 0.6774$, are consistent with recent Planck values (\citealt{2016A&A...594A..13P}).

The simulation was run using the quasi-Lagrangian \texttt{AREPO} code (\citealt{10.1111/j.1365-2966.2009.15715.x}) to follow the coupled dynamics of DM and gas cells. Halos are identified in the simulation using a friends-of-friends group finder algorithm, with a linking length of $0.2$ in units of the mean inter-particle separation (\citealt{1985ApJ...292..371D}). Within these groups, the TNG300 further utilizes the \texttt{SUBFIND} algorithm (\citealt{2001MNRAS.328..726S}) to extract the hierarchical substructure to identify subhalos/galaxies. Subhalos are defined as gravitationally bound, locally over-dense particle groups. Each halo includes a central galaxy, located at the minimum gravitational potential energy of the primary subhalo of the group, and satellite galaxies associated with the subhalos with non-zero stellar mass.
 
In this paper, we focus on the galaxy and halo sample at $z = 0.0$, and use the $r$-band absolute magnitude for the galaxy luminosity, which is based on the summed-up luminosities of all the stellar particles of the group.

\subsection{Defining Environment}
\label{subsec:env}

While previous theoretical studies of galaxy assembly bias mainly used the density field of the dark matter particles as the large-scale environment of the halo (e.g., \citealt{2018ApJ...853...84Z,2021MNRAS.502.3242X}), it is quite challenging to measure it directly from the observations (cf. \citealt{Burchett2020}). In order to facilitate potential future applications of our new measure, we choose to estimate the environment from the galaxy distribution (e.g., \citealt{10.1093/mnras/stu1886,2018MNRAS.476..741D, 2020MNRAS.496.5463A}). We define the large-scale environment of the halos as the galaxy over/under-density estimated within a sphere of radius $r = 8 h^{-1}\text{Mpc}$ centered on the halos' central galaxies, equivalent to smoothing the galaxy density field with a sphere of this radius. The same environment is then assumed for all galaxies, central and satellites, in a given halo. Given the large scale over which the environment is measured, we do not anticipate any significant change if the environment was measured separately for each galaxy in the halo, as discussed further below.  Similarly, resorting to the galaxies' redshift-space positions (as available in observations) is also expected to only make a negligible difference. 

We use a magnitude-limited galaxy sample consisting of all galaxies brighter than $\mathrm{M_r} - 5 \log{h} = -16.0$, which corresponds to a number density of $n = 0.047 h^{3} \mathrm{Mpc}^{-3}$ and includes about $400,000$ galaxies,  to measure the environment.  This magnitude cut is 
motivated by the resolution limit of the simulation, and we have confirmed that our galaxy sample is complete by comparing to the Illustris-TNG100 simulation, which has a higher resolution. We hereafter refer to this galaxy sample as the ``Full sample'' in our analysis. Specifically, we estimate $\delta_8$, the density fluctuation field within $8 h^{-1}\text{Mpc}$ positions of the halos, using the standard definition  $$\delta_8 = \frac{N_g - \bar N_g}{\bar N_g},$$ where $N_g$ is the number of tracer galaxies within a sphere of radius $r = 8 h^{-1}\text{Mpc} $ around each halo's (central galaxy) location, and $\bar N_g $ is the mean value over the whole sample. $N_g$ is measured by computing the number of all galaxy pairs around the central galaxy within the sphere using \texttt{Corrfunc} (\citealt{2020MNRAS.491.3022S}). We do not anticipate our findings to depend on the specific working assumptions made here, in particular since our environment measure is only used later to associate each galaxy to broad environment bins. 

As the environment is only calculated at the halo positions, $N_g \geq 1$, so that the most underdense environment will still have $\delta_8 > -1$, and all galaxies within the halo share the same environment. Again, this is a reasonable estimate since the density smoothing scale, $8 h^{-1}\text{Mpc}$, is significantly larger than the size of any halo.  Considering the intra-halo contribution to the environment, we tested the robustness of our results when discounting all galaxies within a smaller sphere of typical halo size, to obtain a better ``external'' environment measure or, alternatively, explicitly removing all galaxies inside a given halo when estimating its environment. We find the same qualitative trends and similar quantitative measures when simply using $8h^{-1}\text{Mpc}$ spheres, for all the results shown in this paper.

Our choice of the $r = 8 h^{-1}\text{Mpc}$ radius is motivated by the smoothing scale adopted in prior measurements of the environment-dependent LF in observations \citep{2005MNRAS.356.1155C,10.1093/mnras/stu1886,2018MNRAS.476..741D}. Different sphere sizes are discussed in Appendix~B of \citet{2005MNRAS.356.1155C}, concluding that the $8 h^{-1}\text{Mpc}$ scale is a good probe for both the underdense and overdense regions, capturing the essential features while optimizing the statistical signal in the observational sample. Alternative environment measurements are investigated and compared in \cite{2012MNRAS.419.2670M}, finding that the large-scale environment external to a halo is more accurately measured using fixed apertures compared to using nearest neighbor methods and correlates well with the underlying dark matter environment.

The $r = 8 h^{-1}\text{Mpc}$ (top-hat) scale adopted here is also roughly compatible with the $5 h^{-1} \text{Mpc}$ Gaussian smoothing used, e.g., in \citet{2018ApJ...853...84Z}.  Other studies, however, have suggested that the more local environment of the halos may be better suitable for capturing the assembly bias effects \citep{2019MNRAS.482.1900H,2021MNRAS.502.3242X}. Therefore, we also attempted to use the density with a smaller $1.25 h^{-1} \text{Mpc}$ Gaussian smoothing (as adopted in \citealt{2021MNRAS.502.3242X}). However, we find that this smoothing scale is too small and is dominated by the galaxies inside the halos rather than their immediate external environment. Furthermore, it will likely be more sensitive to redshift-space effects in real data. While a measure of this environment directly from the dark matter particles distribution may result in more sensitive results, we conclude that it is not practical when estimating the density from the galaxy distribution. Hence, we chose to proceed with the more robust $r = 8 h^{-1}\text{Mpc}$ smoothing scale for the analysis in this paper.

\subsection{Measuring the Luminosity Function Dependence on Environment}
\label{subsec:LF}

\begin{figure}
\centering	
    \includegraphics[width=\columnwidth]{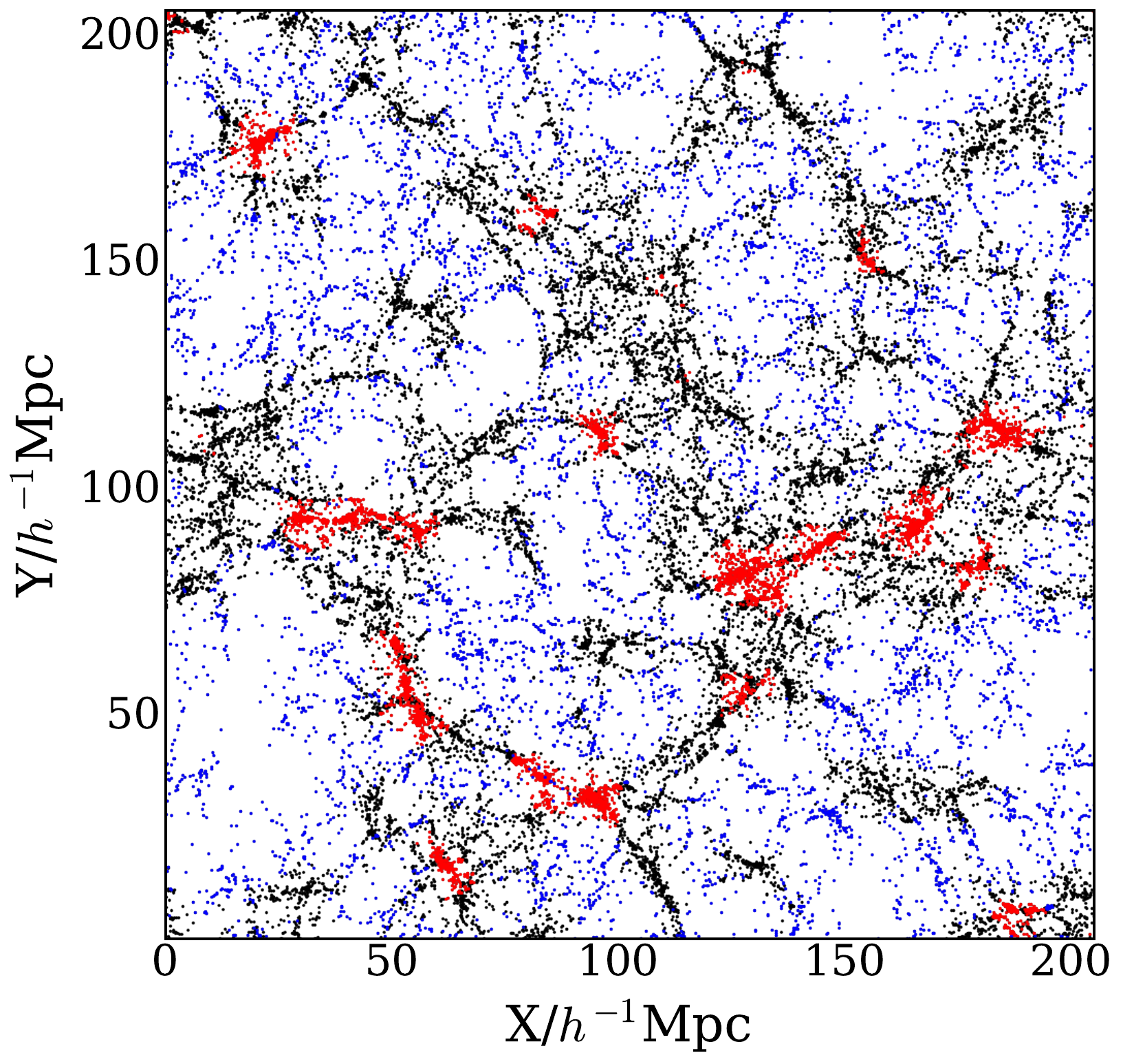}
    \caption{A $205 h^{-1} \text{Mpc} \times 205 h^{-1} \text{Mpc} \times 20 h^{-1} \text{Mpc}$ slice from the TNG300 simulation showing the galaxy spatial distribution. The sample contains all galaxies with $r$-band magnitude brighter than $-16.0$. The red dots denote the $20\%$ of the galaxies in the most dense regions (corresponding to $\delta_8 > 0.92$). The blue dots represent the $20\%$ of the galaxies in the least dense regions ($-1<\delta_8 < -0.44$), while the black dots show the remainder of the galaxies between these two extreme environments. }
    \label{fig:1}
\end{figure}

We use our density measure, $\delta_8$, to split the galaxies into broad environment regimes. We divide the full sample into five bins of roughly equal numbers, each containing around $20\%$ of the galaxies, ranked by their $\delta_8$ values. Note that the exact cuts used for our bins are somewhat arbitrary, and different choices can be made without changing our general conclusions. We tested different definitions of the density bins and comment on that below. Our specific choice was simply intended to maintain a similar signal to noise ratio for all density bins, while spanning the full range of densities. The corresponding density thresholds for our specific choice and the exact fraction of galaxies in each are provided in the first three columns of Table~\ref{tab:f_g} (the remaining columns are discussed later in the text). Figure~\ref{fig:1} shows the spatial distribution of the galaxies in a $20 h^{-1} \text{Mpc}$-wide slice from the TNG300 simulation, highlighting the different environment regions. Specifically, galaxies in the densest environment bin (corresponding to $\delta_8 > 0.92$) are shown in red, while galaxies in the least dense bin ($-1.00 < \delta_8 < -0.44$) are shown in blue. As evident from the plot, galaxies in the most overdense and least dense regions occupy disjoint parts of the cosmic web of structure. We can also qualitatively see that the galaxies in the denser regions are more strongly clustered than those in the underdense regions, as expected.

\begin{deluxetable}{cc|ccc|ccc}
\tablewidth{0pt}
\tablecaption{The $\delta_8$ thresholds for the five density bins and the corresponding galaxy fractions ($f_{\delta}$) listed as percentages, for the full sample and when dividing into red and blue galaxies. These percentages are specified for both the original TNG300 sample (middle section of table) and for the shuffled galaxy sample (right section).
  \label{tab:f_g}}
\tablehead{
  \colhead{$\delta_8^{\text{min}}$} & \colhead{$\delta_8^{\text{max}}$} & \colhead{$f_{\delta}^{\text{Full}}$} & \colhead{$f_{\delta}^{\text{Red}}$} & \colhead{$f_{\delta}^{\text{Blue}}$} & \colhead{$f_{\delta,{\rm Sh}}^{\text{Full}}$} & \colhead{$f_{\delta,{\rm Sh}}^{\text{Red}}$} & \colhead{$f_{\delta,{\rm Sh}}^{\text{Blue}}$} }
\startdata 
0.92 & $\infty$ & 20.0 & 38.2 & 13.9 &  18.5 & 33.2 & 13.6 \\
0.26 & 0.92 & 19.9 & 23.9 & 18.6 &  19.2 & 23.4 & 17.9 \\
-0.12 & 0.26 & 20.2 & 18.4 & 20.8 &  20.1 & 19.3 & 20.3 \\
-0.44 & -0.12 & 19.8 & 12.0 & 22.4 &  20.3 & 14.0 & 22.3 \\
-1.00 & -0.44 & 20.1 & 7.4  & 24.3 &  21.8 & 10.0 & 25.8 \\
\enddata
\end{deluxetable}

While assembly bias has often been probed using the clustering of galaxies, here we aim to study it using an even more fundamental statistic, the galaxy luminosity function. The LF, $\Phi$, is defined as the number density of galaxies per luminosity interval, as a function of luminosity. In our work, we use $r$-band absolute magnitude $\mathrm{M_r} - 5\log h$ for the galaxy luminosity and measure the LF in the TNG300 simulation box with a volume of $205^3 h^{-3} \text{Mpc}^3$. In order to study the environment dependence, we split the galaxy sample into the aforementioned density bins defined by $\delta_8$ values, and measure the LF, $\Phi^{\delta}$, independently in each environment bin. The differences in the shape and slope of the LFs between the density bins show the environment dependence of the LF.

We note that earlier measures of the environment-dependent LF normalized the LF in different ways to account for the effective volume covered by a given density bin \citep{2005MNRAS.356.1155C,10.1093/mnras/stu1886,2018MNRAS.476..741D}. These normalizations adjust the amplitude of the measured LFs across different density bins, resulting in a monotonic trend where the LF amplitude (number of galaxies per effective volume) increases with density. Here, we are interested specifically in the contribution of assembly bias to the LF dependence, so we mainly focus on measuring the ratios of the LFs in different density bins relative to samples sharing the same environment that do not include assembly bias (\S~\ref{subsec:shuffle}). We thus choose not to normalize the LFs, since that factor will cancel out when examining the ratio.

We compute error bars on our measurements of the LF in different density bins (and all derived statistics) using jackknife resampling. To do so, we divide the full simulation volume into $27$ separate cubic subvolumes, by slicing each spatial axis into $3$ equal parts. Each jackknife realization is comprised of the full simulation box excluding in turn one of the individual subvolumes. We repeat our analysis for each jackknife realization in turn, where each one measures, in a sense, the variance arising due to the excluded subvolume. The error bars are then estimated by summing up the total dispersion of all jackknife realization (see, e.g., \citealt{1993stp..book.....L},  \citealt{2002ApJ...571..172Z}, \citealt{2009MNRAS.396...19N}). We have verified that our results are insensitive to the number of subvolumes used. 

\subsection{The Shuffling Mechanism}
\label{subsec:shuffle}

To measure the contribution of assembly bias to the environmental dependence of the galaxy LF, we need to generate a control galaxy sample without any assembly bias to compare with the original sample. To achieve this, we utilize the ``shuffling method'' which is common practice for measuring the impact of GAB on galaxy clustering (e.g., \citealt{2007MNRAS.374.1303C,2018ApJ...853...84Z}). The basic idea of the shuffling is to randomly reassign the full galaxy content of the halos among all halos of similar mass, to remove any dependence on environment or halo properties other than halo mass. The shuffling eliminates a dependence of the galaxy population on any inherent properties of their host halos other than mass. Similarly, any dependence on the large-scale environment of the halos is erased, other than that arising from the halo mass function dependence on environment.

Following the methodology of \citet{2007MNRAS.374.1303C}, we select halos in $0.1$ dex bins of halo mass and randomly shuffle their central galaxies among all halos within the same mass bin. The satellite galaxies are moved along with their original central galaxy, preserving their distribution around it. Standard applications of the shuffling method physically change the positions of all galaxies accordingly, which is crucial when analyzing the impact on galaxy clustering. In our case here, we are measuring the LF dependence on environment and not the correlation function, so the key aspect is the random reassignment of the galaxies to a ``new'' halo environment in the same mass bin. Thus the shuffled sample can, alternatively, simply be generated by randomly reassigning the $\delta_8$ environment measure of each halo among halos of the same mass (without changing the position of the central and satellite galaxies within the halos). We verified that these two methods are equivalent and produce identical results for the environment dependence of the LF.  We also confirmed that the results we present below are insensitive to our specific choice of bin size in halo mass, and that using smaller bins produce only negligible differences. As standard practice, we perform the shuffling algorithm $20$ times with different random seeds, and use the averaged results of these samples throughout the analyses in this paper. The uncertainty from the $20$ random shuffling runs is negligible compared to the jackknife errors introduced in \S~\ref{subsec:LF}. 

While the shuffling changes the galaxy population among halos of the same mass, the underlying dark matter density distribution and other halo properties are not changed by the shuffling procedure. Therefore, we use the same $\delta_8$ measures associated with each halo throughout our work, and only change the correspondence between the environment and galaxy sample when shuffling. However, the environment of a shuffled galaxy sample will, in fact, be slightly different if measured with the shuffled galaxy positions. This arises since the shuffling process, which reassigns the locations of central and satellite galaxies among halos of the same mass, aims to remove assembly bias effects which are themselves correlated to environment. This is a limitation of using the galaxy positions as a proxy for the underlying dark matter density field. While it does not impact our theoretical study of assembly bias in the environment dependence of the LF, it may have implications for practical attempts to measure it.  We discuss this further in Appendix~\ref{sec:contam}. 

Hereafter, in an analogous fashion to estimates of GAB from clustering, we define the assembly bias contribution to the environment dependence of the LF as the ratio between the LF of the original sample to that of the shuffled galaxy sample, computed in a certain environment, $\Phi^{\delta}/\Phi^{\delta}_{\mathrm{sh}}$.

\section{Environmental Dependence of the Luminosity Function}
\label{sec:result1}

\begin{figure*}
\centering
    \includegraphics[width=\textwidth]{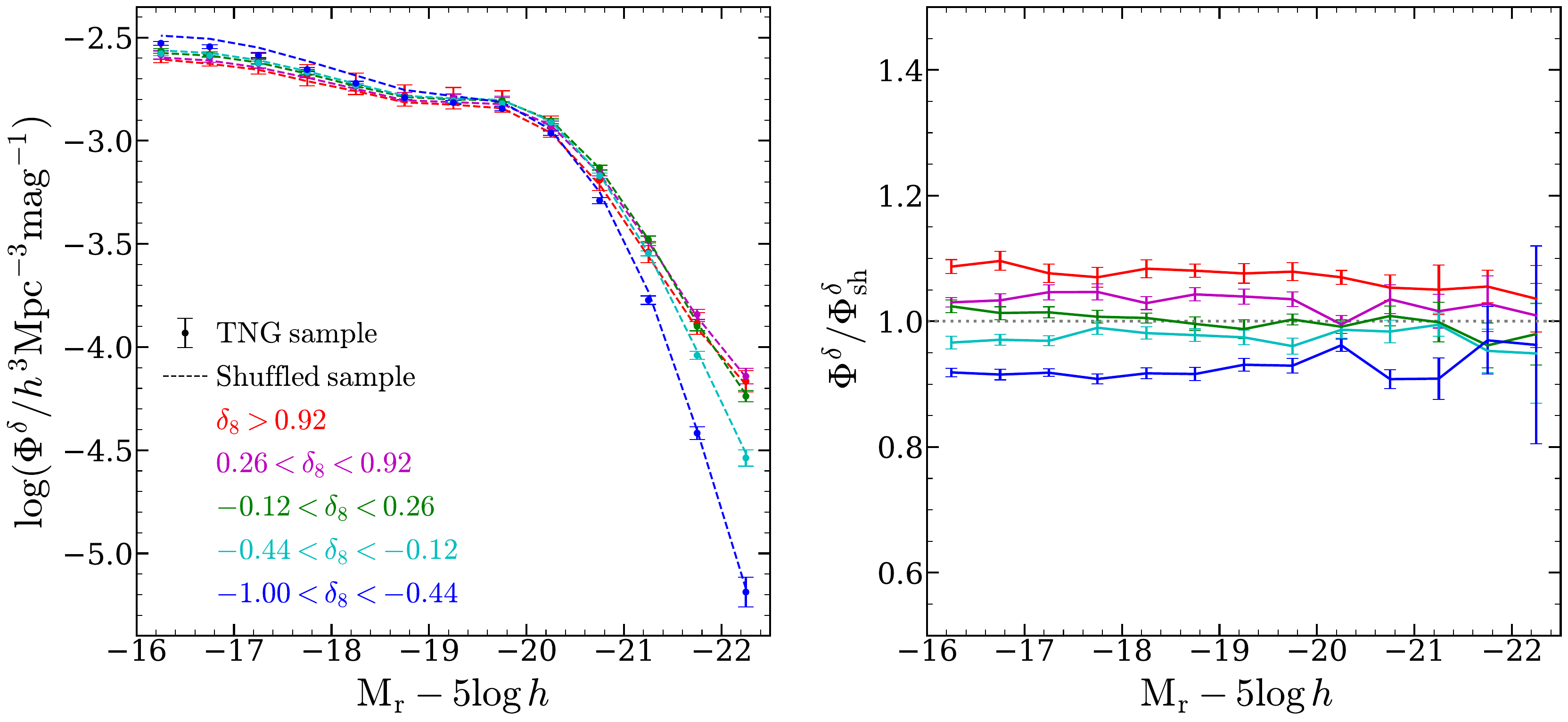}
    \caption{Luminosity functions for different environment for the TNG300 galaxy sample and the assembly bias signatures. Left panel: The LFs of the original sample (solid symbols) and the shuffled galaxy sample (dashed lines), shown separately for our five $\delta_8$ density bins (as labeled). Error bars are obtained from $27$ jackknife realizations (shown for the original measurements only, for clarity). The LFs of the shuffled sample are averaged over $20$ random shuffling runs, to reduce uncertainties. Right panel: Ratios of the LFs of the original and shuffled galaxy samples for each density bin (color-coded the same), exhibiting the impact of assembly bias. The error bars shown are the jackknife uncertainties obtained for the ratios. The dotted gray line indicates the null signal level, a ratio value of $1$.}
    \label{fig:2}
\end{figure*}

In this section, we present our main results for the LF dependence on environment for the full sample and study the assembly bias impact on it. 
The left panel of Figure~\ref{fig:2} shows the luminosity functions measured for the TNG300 galaxy sample in the $5$ density bins, as labeled.  The colored solid symbols represent the number density of galaxies in each magnitude bin and the error bars are the uncertainties of each measurement estimated from the jackknife resampling. Each density bin contains approximately $20\%$ of the full magnitude-limited galaxy sample (see \S~\ref{subsec:LF}). The dashed lines shown are the LFs corresponding to the shuffled samples, color-coded and split by the same density cuts as the original sample. The shuffled samples are obtained as described in \S~\ref{subsec:shuffle}, and we average the results from $20$ randomly shuffled samples. To highlight the differences between the original and shuffled samples, the right panel of Fig.~\ref{fig:2} presents the ratios of the LFs of the original sample and those corresponding to the shuffled samples, calculated separately for each environment bin (solid lines, color coded accordingly). The error bars represent the uncertainties of each measurements estimated from the jackknife resampling, fully propagated to the ratios shown. The uncertainty associated with the $20$ random shuffling runs is not included, as it is negligible compared with the jackknife errors.

We start our analysis by examining the left panel of Fig.~\ref{fig:2}, demonstrating the dependence of LF on environment. We find clear differences between the LFs in different density bins at both the faint end and bright end. Most notably, the LF of galaxies in the least dense regions is steeper at the faint end and has a sharper break of the LF at the bright end.  For brighter luminosities (above the break), we see that there are more galaxies in the denser regions versus the underdense ones.  This trend reverses for galaxies at the faintest luminosities which are more prevalent in the underdense regions. This is consistent with our expectation that brighter galaxies tend to reside in dense environments, whereas faint galaxies reside more in lower-density regions.   We remind the reader that the density bins are chosen here so as to have similar number of galaxies, and we do not employ any additional normalization to account for the effective volume covered, which impacts the amplitudes of the different LFs. Once accounting for this, the trends we find are in agreement with the results of \citet{10.1093/mnras/stu1886} for the environement dependent LF in the GAMA survey.

We now turn our attention to the LF measurements for the shuffled galaxy samples (dashed lines in the left panel of Fig.~\ref{fig:2}), which show the results once removing the effects of assembly bias.  The LFs maintain the qualitative relative trends with environment and magnitude. The overall similar shapes of the LFs in different density bins after the shuffling indicate that the dependence of the LF on environment mostly arises from the dependence of halo mass on environment (see Appendix~\ref{sec:HMF} for more detail). However, there are distinct, and systematic, differences between the LFs of the original and shuffled samples (noted by comparing the solid dots to the dashed lines of the same color).  These differences are due to assembly bias and are the focus of the rest of this paper. 

We get the first hint of these differences by comparing the overall galaxy fraction in each density bin before and after the shuffling. These fractions are provided as well in Table~\ref{tab:f_g}. For the original sample, these fractions are $\sim20\%$ for all bins, by construction. The shuffling changes that, and we find a monotonic trend with density where only $18.5\%$ of the sample in the densest bin and $21.8\%$ in the least dense bin (sixth column of Table~\ref{tab:f_g}). Namely, assembly bias, which is present in the original sample, tends to increase the number of galaxies in dense environments and decrease the number of galaxies in underdense regions. Hence, the influence of assembly bias changes the galaxy populations in different environments on top of the impact of halo mass. 

The right panel of Fig.~\ref{fig:2} focuses on these differences, showing the ratios of the two LFs, $\Phi^{\delta}/  \Phi_{\mathrm{sh}}^{\delta}$, for all density bins.  We find distinct signatures of assembly bias, as shown by the solid lines, indicated by deviations from a ratio of $1$. The values of these ratios increase monotonically with the $\delta_8$ density, and are largely independent of the galaxy luminosity in each density bin. This is a key result of our paper, demonstrating the impact of assembly bias on the environment-dependent LF. In the denser environments, the amplitude of the LF of the original sample is significantly higher than that of the shuffled sample (sans assembly bias), corresponding to a ratio larger than $1$. This indicates that assembly bias tends to increase the prevalence of galaxies in denser environment. The reverse holds for the underdense environments, where the original LF has a lower amplitude than for the shuffled sample (ratio smaller than $1$), while for the intermediate density bin (around $\delta_8 \sim 0$), the two LFs are roughly comparable.

These results are consistent with the analysis and understanding that emerged by \citet{2018ApJ...853...84Z}. As discussed in the latter paper, for fixed halo mass, galaxies above a given stellar mass or luminosity threshold tend to more readily populate halos in denser environments (or earlier formation time), reflected in the trends exhibited in the occupancy variation, scatter of the stellar mass-halo mass relation and galaxy assembly bias. Here we are probing this effect more directly, by examining the LF (number of galaxies) as a function of environment. We find that assembly bias gives rise to an additional clear tendency of denser regions to be more occupied by galaxies (relative to a shuffled galaxy sample with no assembly bias effects), and a relative paucity of galaxies in underdense regions. We stress that these are secondary effects on top of the fundamental dependence of the halo mass on environment. The magnitude of the effect for the extreme bins in over/under-density is nearly
$10\%$ in the amplitude of the LF.  The signal is statistically very significant over most of the luminosity range, given the levels of the measurement errors, with the exception of the brightest magnitudes where the errors increase due to the paucity of such galaxies (see left panel of Fig.~\ref{fig:2}).

Our results are fairly robust to the specific density bins used. We tested variations of it using bins of $\sim 10\%$ of the galaxies each, or the specific density cuts assumed in \citet{10.1093/mnras/stu1886}. We obtained identical qualitative trends and small variations in the measured values. For the $10\%$ bins, the extreme density bins exhibit a slightly larger effect, but still of the order of $10\%$ and with slightly larger error bars, as expected.  We find the $20\%$ density bins as the best choice in producing both distinct and stable signals, and stick with this choice for the remainder of our analysis. Our results also remain largely unchanged when defining the $\delta_8$ environment from the shuffled galaxy positions as the proxy for the underlying density and using it consistently for both the original and shuffled LF calculations. 
However, using the original galaxies environment for the original sample LF and the shuffled galaxies environment for the shuffled sample LF (with the same density bin thresholds) can introduce systematics that change the measured signal. See Appendix~\ref{sec:contam} for more details. 

\begin{figure}
\centering
    \includegraphics[width=\columnwidth]{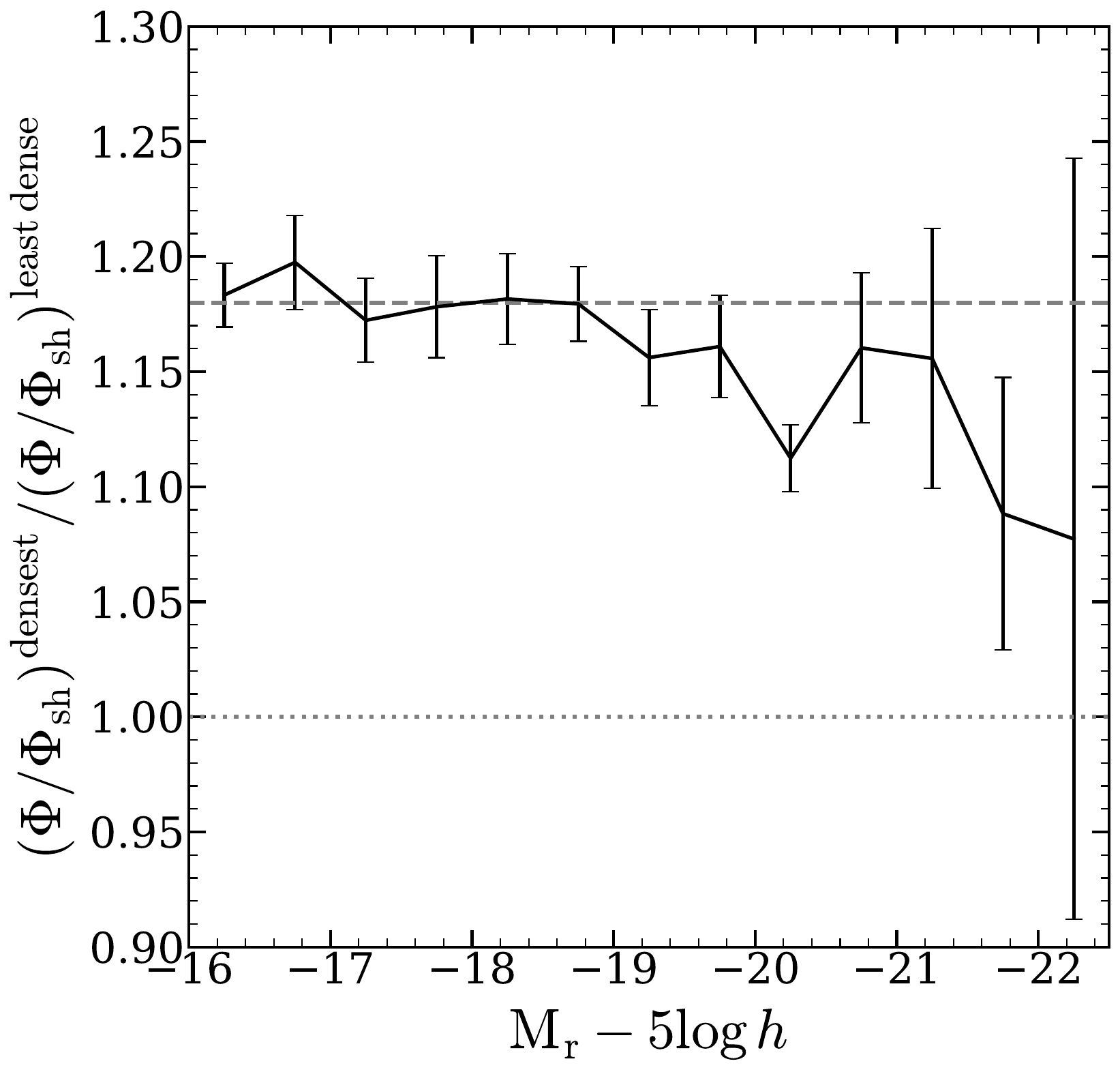}
    \caption{The ratio between the assembly bias signatures of the most dense and least dense environments. The black solid line shows the ratio of $\Phi^{\delta}/\Phi_{\mathrm{sh}}^{\delta}$ for the $20\%$ densest environment bin to that of the least-dense bin. The dotted gray line indicates the null signal level, namely a ratio of $1$. The dashed gray line at $1.18$ shows the average level of the signal over $-16.0 > \mathrm{M_r} - 5 \log h > -18.5$, for which the measurements are roughly constant. The error bars denote jackknife uncertainties.}
    \label{fig:3}
\end{figure}

To further hone in on the assembly bias signal, we show in Figure~\ref{fig:3} the ratio between the $\Phi^{\delta}/\Phi_{\mathrm{sh}}^{\delta}$ values of the ($20\%$) most and least dense bins. The ratio between the two extreme environments naturally shows a stronger signal of around $18\%$ for galaxies fainter than a magnitude of $-18.5$, where the signal is roughly constant. We note that the error bars on the ratio of $\Phi^{\delta}/\Phi_{\mathrm{sh}}^{\delta}$ for the most dense and least dense bins are somewhat larger than the uncertainties of the two individual density bins, since the variations of the LFs in the most and least dense regions among the different jackknife realizations are anti-correlated. Still, the signal is highly significant over that range given the level of errors. The measured ratio decreases to about $10\%$ at the brightest end, before the error bars become too large.  The decrease of the signal at the bright end agrees with the findings of \citet{2007MNRAS.374.1303C} and \citet{2018ApJ...853...84Z} that fainter galaxies exhibit stronger galaxy assembly bias effects.

\section{Results for Color-Selected Samples}
\label{sec:result2}

To investigate whether the assembly bias effects in the environmental dependence of the LF vary for different types of galaxies, we split the sample by $(g-r)$  color and repeat the previous measurements on red and blue galaxies separately. In \S~\ref{subsec:color}, we motivate splitting our galaxy sample by color and characterize the subsamples. Section~\ref{subsec:GAB_color} then presents our main results for the impact of assembly bias on the environment-dependent LFs for the red and blue galaxies separately.

\subsection{Galaxy Colors}
\label{subsec:color}

Galaxy colors can serve as a valuable indicator of recent star formation history (e.g., \citealt{10.1111/j.1365-2966.2009.15512.x}, \citealt{Maller_2009}, \citealt{10.1111/j.1365-2966.2012.21188.x} ), and are also correlated with the galaxy environment along with other properties such as metallicity and gas content (\citealt{1983AJ.....88..483K}, \citealt{10.1046/j.1365-8711.2002.05558.x}, \citealt{2003ApJ...584..210G}, \citealt{2003ApJ...585L...5H}, \citealt{2005ApJ...629..143B}). 

\citet{2007MNRAS.374.1303C} investigated GAB in galaxy clustering as a function of luminosity and color in the Millennium Simulation, generally finding stronger effects for faint red galaxies. They demonstrate that the color of the central galaxy in a halo of a given mass depends significantly on the halo's environment, finding a particularly strong effect (large bias) for red central galaxies. This is attributed to these galaxies being either in massive halos where cooling and star formation are quenched by AGN feedback or in lower mass splashback halos that are stripped of their hot gas. Both cases are associated with denser environments and hence larger clustering amplitude (see also \citealt{2003ApJ...585L...5H}). In contrast, blue central galaxies are associated with more isolated and/or lower mass halos that have ongoing star formation.

\begin{figure}
\centering
    \includegraphics[width=\columnwidth]{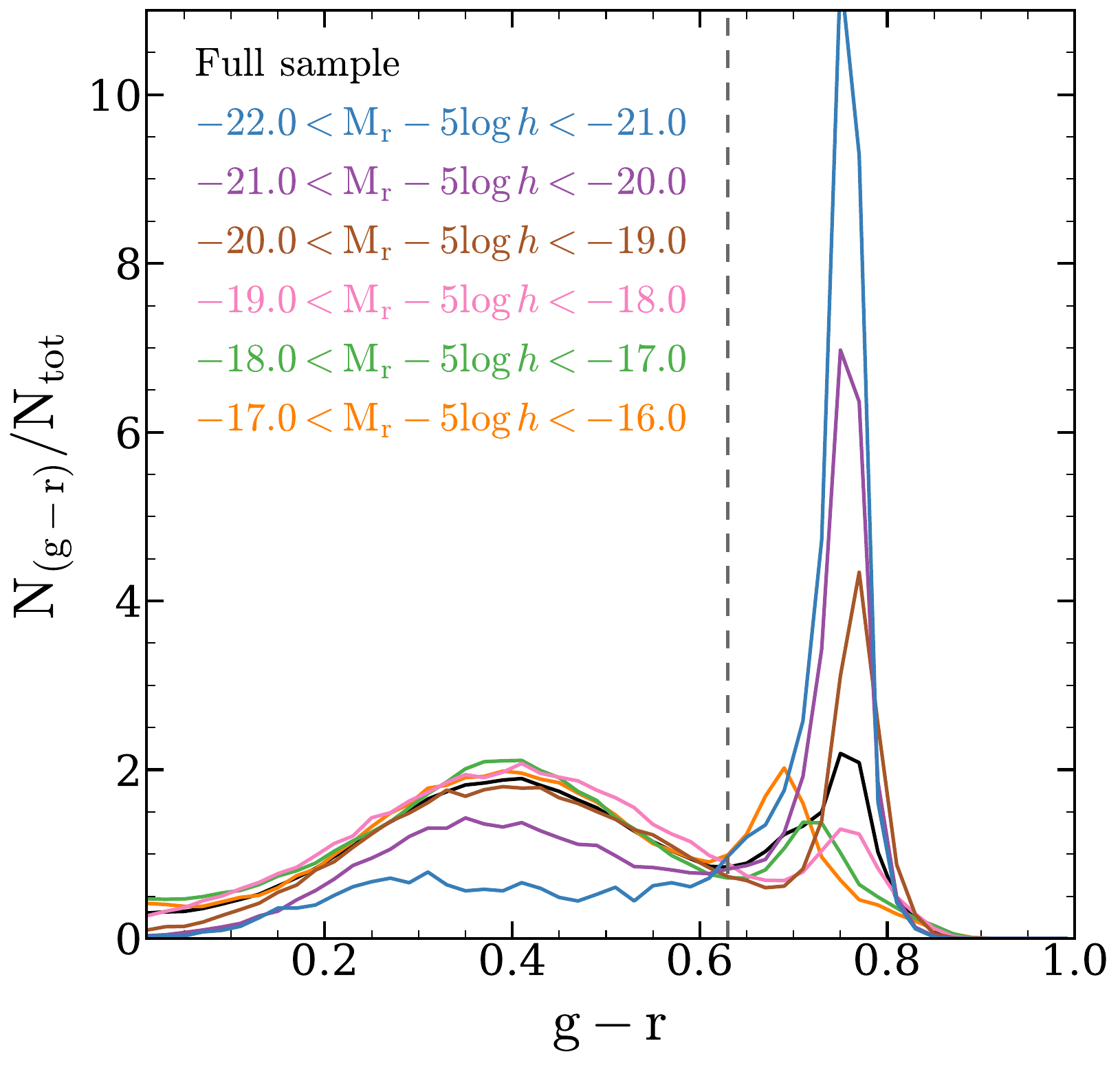}
    \caption{Distribution of rest frame $(g-r)$ color for our full sample from the TNG300 simulation (solid black line) and separately for different ranges of $r$-band absolute magnitude, as labeled. The vertical dashed line denotes the color cut adopted, at $(g-r) = 0.63$, for splitting the galaxy sample into red and blue populations. }
    \label{fig:4}
\end{figure}

Motivated by this, we split the galaxy sample by color to further investigate the manifestation of assembly bias on the LF in different environments. Galaxy colors are defined by the $(g-r)$ rest-frame color obtained directly from the $r$-band and $g$-band absolute magnitudes. The simulated $(g-r)$ colors of TNG galaxies at low redshift are in good agreement with observational data from the SDSS \citep{2018MNRAS.475..624N}. Figure~\ref{fig:4} shows the distribution of the $(g-r)$ color for the full galaxy sample (solid black line) and galaxies in different magnitude ranges (solid lines in different colors) from the TNG300. Following the clear bimodality exhibited in the plot, we choose to divide the full sample into a blue and red population using a color cut of $(g-r) = 0.63$, as indicated by the dashed vertical line. This value corresponds to the `minimum' in the $(g-r)$ color distribution of the full sample, and also shows a good split for the galaxies within varying magnitude ranges.  Galaxies with $(g-r)$ values larger than this cut make up our red galaxies population, consisting of about $25\%$ of the full sample, with the remainder making up the population of blue galaxies.

Next, we examine the connection to environment. We use the same five bins defined previously for the full sample.  Table~\ref{tab:f_g} shows the division of each color subpopulation among these different environments. While the full sample was split roughly equally among them, we find variations for red and blue galaxies.  As expected, red galaxies preferentially reside in the overdense regions and only a smaller fraction of them populate the underdense regions. The opposite is true for the blue galaxies, which tend to populate more the underdense regions. Similar trends were seen in observations (e.g., \citealt{1980ApJ...236..351D}, \citealt{2005MNRAS.356.1155C, 10.1093/mnras/stu1886}). For TNG300, in the least dense environment bin, we find that only $\sim 9\%$ of all galaxies are red and the rest are blue, while in the densest environment bin, the red galaxy fraction increases to $\sim 48 \%$.

\begin{figure}
\centering
    \includegraphics[width=\columnwidth]{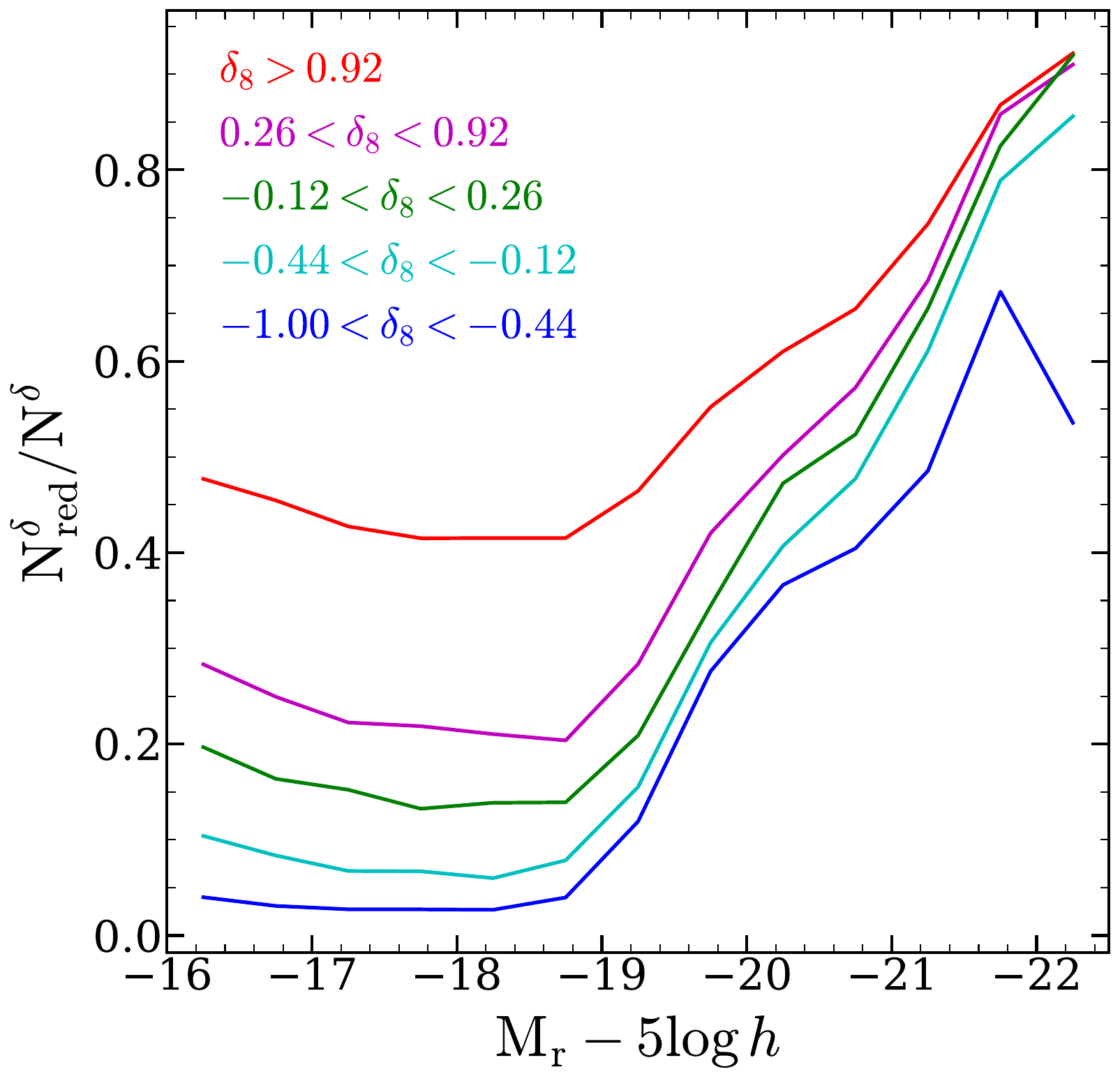}
    \caption{Red galaxy fractions in each of the five density bins as a function of $r$-band absolute magnitude (colored solid lines, as labeled). The values represent the fraction of red galaxies out of the total in each combined density and magnitude bin.}
    \label{fig:5}
\end{figure}

Figure~\ref{fig:5} shows in more detail how the fractions of red galaxies in different environments vary with galaxy luminosity.  We note once again the overall prevalence of red galaxies in dense environments, as well as the tendency of brighter galaxies to have red colors, largely irrespective of environment. This latter general trend is well-established and is also evident in Fig.~\ref{fig:4}. For galaxies with fainter magnitudes, we see in Fig.~\ref{fig:5} that the fraction of red galaxies is very sensitive to the environment, and increases strongly with density. This suggests a stronger correlation between the red galaxy population and environment at the faint end. These insights will be useful for us as we analyze next the assembly bias trends in the environment-dependent LFs for galaxy samples split by color.

\subsection{Environmental Dependence of the Luminosity Function for Red and Blue galaxies} \label{subsec:GAB_color}

\begin{figure*}
\centering
	\includegraphics[width=\textwidth]{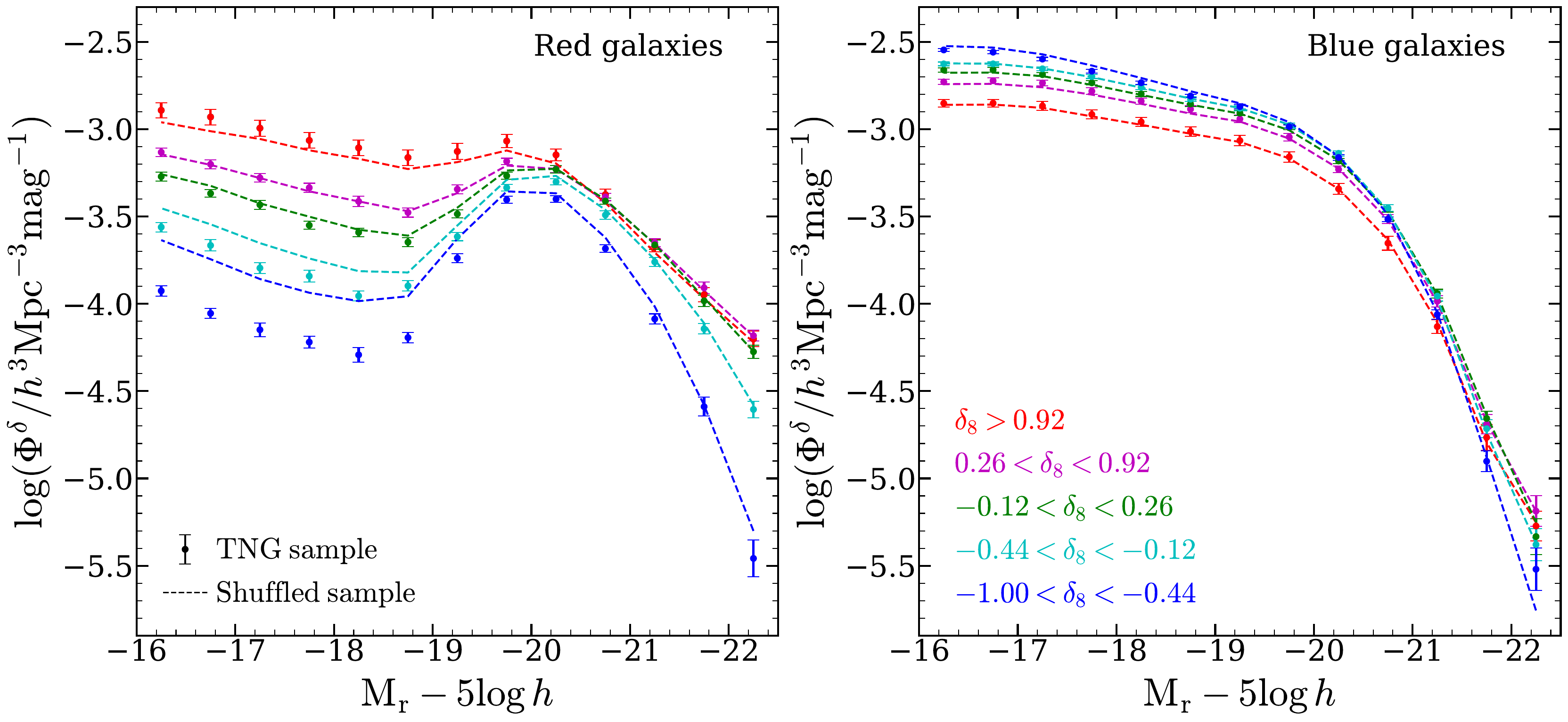}
    \caption{ Luminosity functions for the TNG300 red galaxy sample (left panel) and blue galaxies (right panel) in the five different environment bins, as labeled. As in Fig.~\ref{fig:2}, the symbols are the measurements for the original sample, and the corresponding dashed lines are the measurements for the shuffled sample for each environment bin. The shuffled results are averaged from $20$ random shuffling runs. Errors in each panel are derived from $27$ jackknife samples. }
    \label{fig:6}
\end{figure*}

Figure~\ref{fig:6} shows the luminosity function measured separately for the red galaxies (left panel) and the blue galaxies (right panel) from the TNG300 simulation. As in Fig.~\ref{fig:2}, we measure the LFs in the same five $\delta_8$ density bins, shown as color-coded symbols with jackknife error bars for the original samples and as dashed lines for the shuffled samples (averaged over 20 shuffling runs). We stress that the shuffling process is unchanged and is performed before splitting the galaxy sample by color. We also note that $\delta_8$ is computed as before from the full galaxy sample, serving as a proxy for the underlying dark matter density field. 

As shown in Fig.~\ref{fig:6}, there are distinct differences between the LFs for red and blue galaxies, for all environment bins. While the LFs for the blue galaxies largely resemble a Schechter-function like shape, similar to the LF for the full galaxy sample, the LFs for the red galaxies have a non-monotonic behavior with an upturn at intermediate luminosities, before the dropoff at the bright end. Our results are in general agreement with previous measurements of the LF for color or type-selected samples in observations (e.g., \citealt{2005MNRAS.356.1155C, 2010ApJ...721..193P, 10.1111/j.1365-2966.2011.20111.x, 2013MNRAS.434.3469D, 10.1093/mnras/stu1886}).  This feature in the red galaxies LF is enhanced in underdense regions due to the paucity of faint red galaxies (as shown in Fig.~\ref{fig:5}) and is likely related to the prominence of bright red galaxies, mostly as central galaxies in massive halos. We refer the reader to \citet{2010ApJ...721..193P} for a discussion of the physical processes, related to mass and environmental quenching and merging of passive galaxies, giving rise to this shape and a proposed two-component Schechter-function model for it.

The red and blue LFs also exhibit clear differences in the overall change of amplitude with density,  with the amplitude increasing with density for the red galaxies and decreasing with increased density for the blue galaxies. This again reflects the dominance of blue galaxies in the underdense regions and the prominence of red galaxies in dense regions (see Table~\ref{tab:f_g}). As expected, the differences with environment are most evident for the red galaxies at the faint end of the LF (left panel of Fig.~\ref{fig:6}).  These trends are further amplified by assembly bias effects, as illustrated by comparing the symbols (original sample) to the dashed lines corresponding to the shuffled samples (for which assembly bias is removed by construction). We see that, similar to before, assembly bias effects act to increase the population of red galaxies in dense regions and decrease their numbers in underdense regions. These trends are similar to but significantly stronger than those seen previously for the full sample. In contrast, for the blue galaxies, the variations of the LF with environment are smaller overall. Assembly bias trends are still present, but to a much lesser extent.

\begin{figure*}
\centering
	\includegraphics[width=\textwidth]{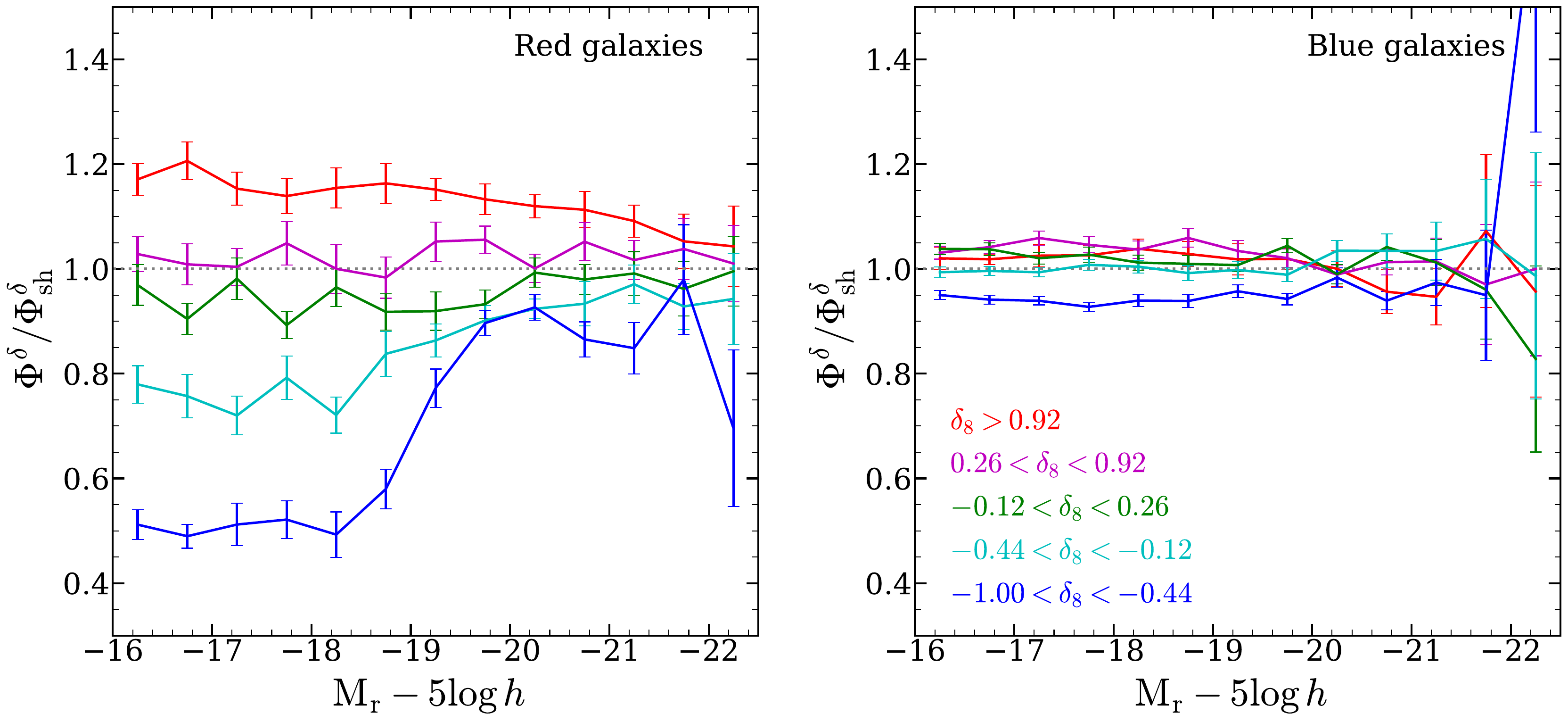}
    \caption{Assembly bias signals in the environment dependence of the luminosity function, computed separately for the red galaxies  (left panel) and blue galaxies (right panel). Colored lines in each panel are the ratios between the original LFs and those of the shuffled galaxy samples in the density bins, as labeled. The shuffled results are averaged from $20$ random shuffling runs. The errors in each panel are jackknife errors.}
    \label{fig:7}
\end{figure*}

The assembly bias effects are highlighted in Figure~\ref{fig:7}, where we extract the signal by calculating the ratios of the original LFs and the ones corresponding to the shuffled galaxy samples in each environment bin, separately for red and blue galaxies. For the blue galaxies, the right panel of Fig.~\ref{fig:7} presents relatively weak trends which exhibit less clear dependence on density, and are roughly independent of luminosity. The change in the LF due to assembly bias is at most $\sim 5\%$, about half the size of the effect for the full sample (Fig.~\ref{fig:2}). Meanwhile, in the left panel of Fig.~\ref{fig:7}, we find substantially stronger signals for the red galaxies, especially at the faint end. Assembly bias accounts for about a $20\%$ increase in the number of faint red galaxies in the densest environment, and a $50\%$ decrease in the least-dense environment! In contrast to the luminosity-independent signals for the full sample and the blue galaxies, the signatures for the red sample show a notable distinction between the faint and bright galaxies. The strong assembly bias signal for the faint-end red galaxies is roughly constant in amplitude over $-16.0> \mathrm{M_r} - 5 \log h> -18.5$, for each density bin. For brighter red galaxies, the signal decreases to a $\sim 10\%$ level (for the extreme environment bins), comparable to the results for the full sample.

\begin{figure*}[t]
  \includegraphics[width=0.95\textwidth]{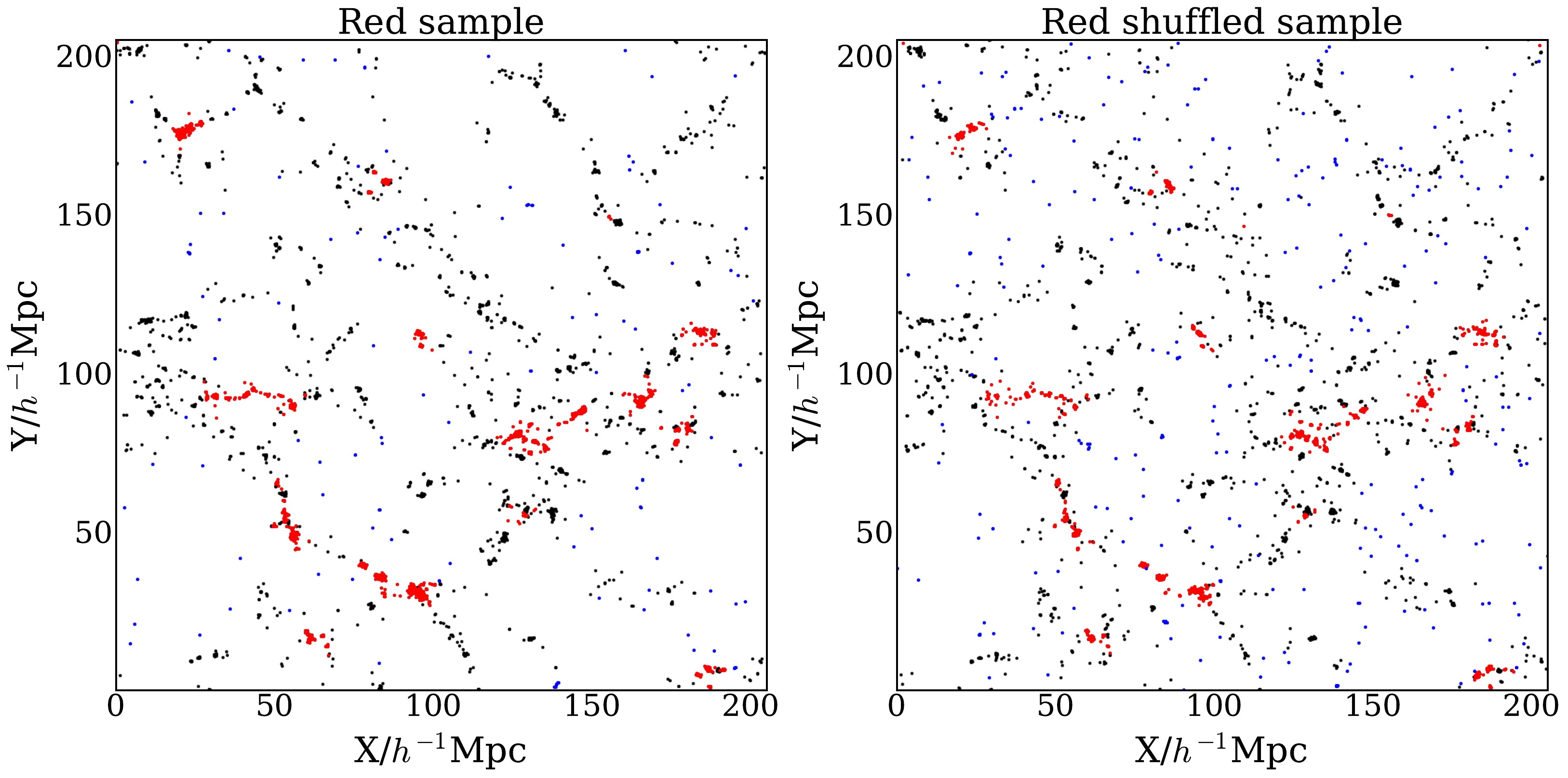}
\caption{The spatial distribution of faint red galaxies in a $205 h^{-1} \text{Mpc} \times 205 h^{-1} \text{Mpc} \times 20 h^{-1} \text{Mpc}$ slice of the TNG300 simulation. The figures include only the red galaxies with $r$-band magnitude $-16.0> \mathrm{M_r} - 5 \log h> -18.5$.  The left panel shows the original positions of the galaxies in the slice, while the right panel shows the faint red galaxies after shuffling the galaxy positions for fixed halo mass (for one realization). Note that the color of the dots does not indicate the galaxy's color, but instead marks the environmental density, consistent with the previous plots. The red dots represent the most dense region ($\delta_8 > 0.92$), the blue dots represent the least dense region ($-1.00 <\delta_8 < -0.44$), and the black dots show the remainder of the galaxies between these two extreme environments. }
    \label{fig:8}
\end{figure*}

The assembly bias signal is, in fact, so strong for faint red galaxies that we can see it ``by eye''.  Figure~\ref{fig:8} shows the distribution of the red galaxies with $r$-band magnitude  $-16.0> \mathrm{M_r} - 5 \log h> -18.5$ in a slice from the TNG300 simulation (the same one as in Fig.~\ref{fig:1}). Their original positions are shown on the left-hand side while the right panel shows their distribution in one realization of the shuffling.  Both slices contain a similar number of galaxies. In both panels, the red dots mark the faint red galaxies in the densest environment bin with $\delta_8> 0.92$ and blue dots correspond to the faint red galaxies in the least dense bin with $-1.00 < \delta_8 < -0.44$. The defined locations of these regions do not change for the original slice and the shuffled one. However, in the shuffled slice, the galaxies are randomly reassigned to halos of the same mass, irrespective of any other halo property or the environment at large.

When comparing the two panels in Fig.~\ref{fig:8}, the overall impression is that the galaxies in the original sample are more clustered than in the shuffled sample, reflecting GAB, and naturally in each panel the densest regions appear more clustered than the least dense ones. Moreover, comparing the two panels, it is evident that the dense regions are more populated by galaxies (the red dots) in the original sample versus the shuffled one. Assembly bias increases the tendency of galaxies to populate dense environments. Conversely, there are distinctly less galaxies in the least dense regions (blue dots) in the original sample compared to the same regions in the shuffled slice, revealing the opposite effect of further ``emptying'' the voids. These are precisely the assembly bias effects we measure and quantify in the LF. Fig.~\ref{fig:8} thus serves as a good illustration to visualize the impact of assembly bias on the environment-dependent LF.

For completeness, we show in Appendix~\ref{sec:slice} the spatial distribution of all (red and blue) galaxies with $r$-band magnitude  $-16.0> \mathrm{M_r} - 5 \log h> -18.5$ in the same slice, as well as the distribution of just the blue galaxies. For these, one cannot discern visually the assembly bias effects when comparing to the shuffled samples, as the LF differences are much smaller (as shown in the right panels of Fig.~\ref{fig:2} and Fig.~\ref{fig:7}). What is apparent is the larger number density of galaxies overall compared to the paucity of faint red galaxies highlighted in Fig.~\ref{fig:8}. The paucity of the faint red galaxies, which was already discussed in \S\ref{subsec:color} and specifically shown in Fig.~\ref{fig:5}, might make the detection of assembly bias in the LF challenging.

We note that the measured error bars on the LFs and the $\Phi^{\delta}/\Phi^{\delta}_{\mathrm{sh}}$ ratios for red galaxies in figures~\ref{fig:6} and~\ref{fig:7} are larger compared to those of the full and blue samples. This is mostly due to the more clustered and compact spatial distribution of the red galaxies in dense environments, covering a much smaller effective volume (as can be seen in Fig.~\ref{fig:8}), and the relatively small number of faint red galaxies in the underdense regions (Fig.~\ref{fig:5}). Given that, the uncertainties on the red sample measurements are reasonable, and the assembly bias signatures are still highly statistically significant. We checked the robustness of the results by repeating the measurements with different density bin choices, e.g., when using density bins with $20\%$ red galaxy fraction, which all produce the same trends. We also studied the impact of switching to the shuffled galaxies environment measure for the shuffled sample LFs. We find that this systematic is relatively weaker for the red galaxies, resulting in a robust measurement of assembly bias (see Appendix~\ref{sec:contam} for more details). This indicates that our approach will still be able to produce reliable measurements from real data even when estimating the environment from a simulated HOD sample.

\begin{figure}
\centering
	\includegraphics[width=\columnwidth]{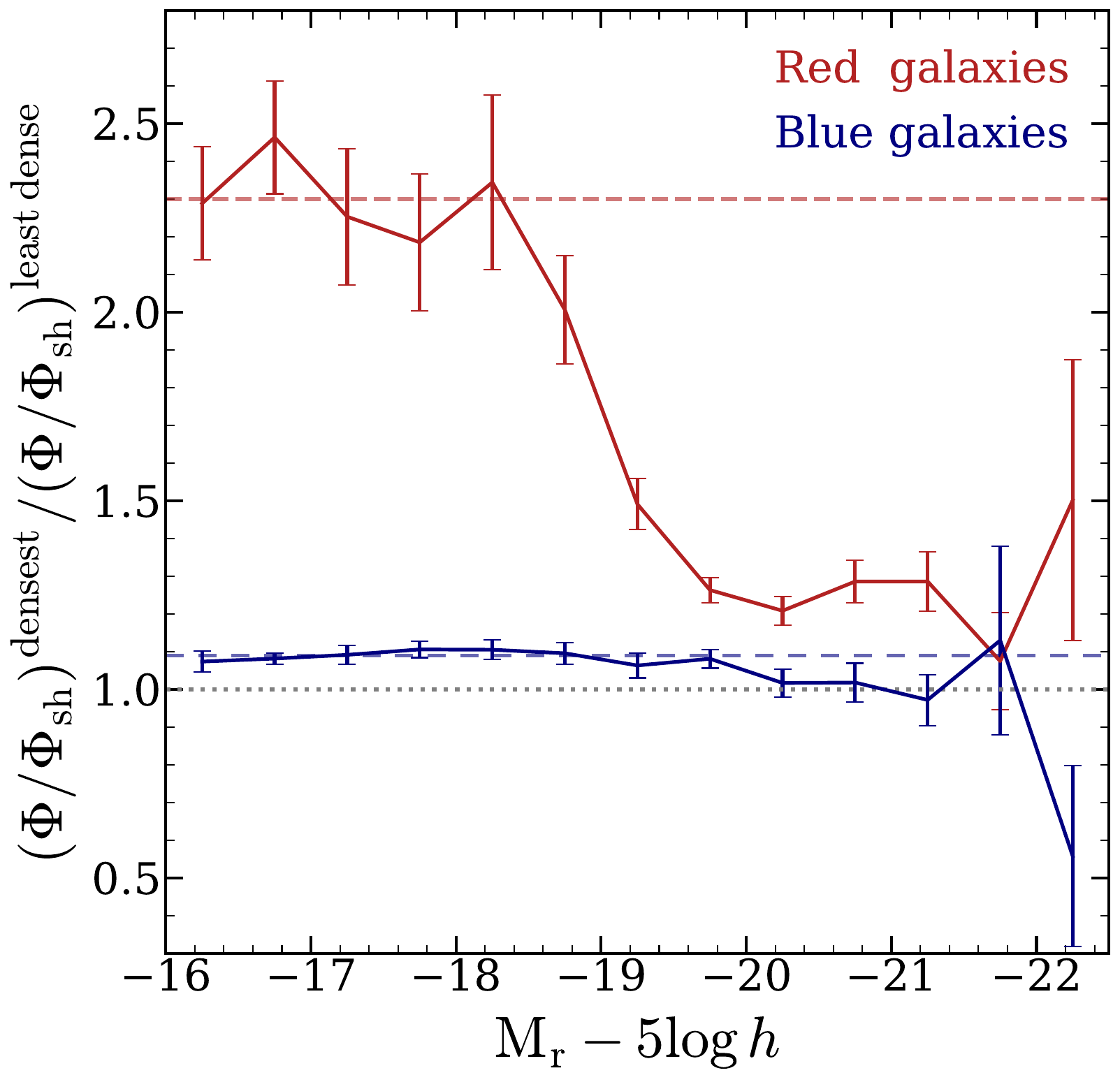}
        \caption{The same as in Fig.~\ref{fig:3} but now for the red and blue galaxies separately. The red and blue solid lines show the ratio of $\Phi^{\delta}/\Phi_{\mathrm{sh}}^{\delta}$ for the densest environment bin to that of the least dense environment, respectively for the red galaxies and blue galaxies. The dotted gray line denote the null signal level. The dashed red and blue lines indicate the signal level averaged over the faint end ($-16.0 > \mathrm{M_r} - 5 \log h > -18.5$) for the red and blue galaxies, over which the ratios are roughly constant. These correspond to values of $2.3$ and $1.1$, for red and blue respectively. The error bars are again jackknife error bars.}
    \label{fig:9}
\end{figure}

It has been well established that red galaxies preferentially populate dense regions and are scarcer in underdense environments. We are seeing here that assembly bias further enhances these trends, in particular for faint red galaxies.
Finally, we show in Figure~\ref{fig:9} the ratio between $\Phi^{\delta}/\Phi_{\mathrm{sh}}^{\delta}$ for the densest environment bin and that for the least-dense environment. For the blue galaxies and for the bright red galaxies we get a signal reminiscent of the results for the full sample.  However, for the faint red galaxies we get a significantly strong signal of $\sim2.3$. This remarkable, {\it larger than a factor of 2}, assembly bias signature is much stronger than assembly bias signals found in previous studies, and may have the potential to be detected in observations. To clarify, we do not necessarily expect the signal in the real universe to be the same as in the TNG300, however, our analysis indicates that this measure is particularly sensitive to assembly bias.

While it is hard to compare our new results to other studies of assembly bias, given the differences in how the signal is measured, the trends we are finding are consistent with previous studies of GAB using galaxy clustering in both numerical models and observations. For example, \citet{2018ApJ...853...84Z}  show that galaxies tend to more readily populate halos in denser environments, leading to the different manifestations of assembly bias.  Similarly, a larger GAB amplitude has been predicted by semi-analytical galaxy formation models for larger number densities (roughly comparable to including fainter galaxies) for both stellar mass selected samples and star-formation rate selected galaxies \citep{2019ApJ...887...17Z,2019MNRAS.484.1133C}. Regarding the color dependence of GAB, \citet{2007MNRAS.374.1303C} have shown that red galaxies have a stronger level of assembly bias compared to blue (or all) galaxies, particularly at faint magnitues, in agreement with our measurements. Similar trends have also recently been found in observational constraints of GAB, with red galaxies from the BOSS galaxy survey, having higher amplitudes than bluer samples, such as the ELGs from DESI 1\% \citep{2023MNRAS.525.3149C,2024arXiv241111830O}.

\section{Conclusion}
\label{sec:concl}

We have utilized the cosmological hydrodynamical simulation TNG300 to study the manifestation of assembly bias in the environment-dependent LF. We measure the LFs in five different bins of large-scale environment and compare to the results when randomly reassigning (shuffling) the galaxy positions among halos of the same mass. The density, $\delta_8$,  is estimated from the galaxies positions using an $8 h^{-1}\text{Mpc}$ top-hat smoothing, and the bins are chosen to include $\sim20\%$ of our galaxy sample. We examine the ratios of the LF measurements, $\Phi^{\delta}$, to those of the shuffled samples, $\Phi_{\mathrm{sh}}^{\delta}$, to explore in detail the impact of assembly bias. We focus on the results for the densest environment bin and the least dense bin, which exhibit the strongest signals, and also explore color-selected samples. We summarize our main results as follows:

\begin{itemize}

\item The environmental dependence of the LF exhibits considerable assembly bias effects, which amplify the tendency of galaxies to reside in dense environments and conversely further devoid underdense regions of galaxies, on top of the primary trends driven by halo mass.

\item The assembly bias signal exhibited in the ratio of the original to shuffled LFs varies smoothly with environment, with $\Phi^{\delta}/\Phi_{\mathrm{sh}}^{\delta}$ increasing monotonically with $\delta_8$. The signatures for the most (least) dense bins are close to a $10\%$ increase (decrease) in the amplitude of the LF across all luminosities, as can be seen in the right panel of Fig.~\ref{fig:2}. 

\item The influence of assembly bias on the environment dependence of LF can be highlighted by examining the ratio of $\Phi^{\delta}/\Phi_{\mathrm{sh}}^{\delta}$ between the most and least dense environments, which produces a significant signal of about $18\%$ for galaxies with $\mathrm{M_r} - 5 \log{h} > -18.5$.

\item When split by color, the shape of the LFs of red and blue galaxies exhibit distinct differences, with the LFs of red galaxies showing clear deviations from a standard Schecter-function like shape and a greater dependence on environment.
  
 \item The LFs for blue galaxies have similar overall shape to those of the full sample, are more prevalent in underdense regions, and exhibit weaker assembly bias effects (at the $5\%$ level).

 \item The red galaxies show very distinct and significantly stronger assembly bias, especially at the faint end. Assembly bias accounts for a $\sim20\%$ increase in the number of galaxies in the densest environment bin, and a striking $50\%$ decrease in the least dense regions, as presented in Fig.~\ref{fig:7}.  These trends decrease to a $\sim10\%$ level at the bright end.

\item We see that assembly bias acts to enhance the prevalence of red galaxies in dense regions and paucity in underdense regions (particularly for faint galaxies), as a global environmental effect, beyond the halo mass function dependence on environment. 

\item The strongest signal is obtained from our approach when comparing the $\Phi^{\delta}/\Phi_{\mathrm{sh}}^{\delta}$ ratio for the most dense bin to the least dense bin resulting in a remarkable larger than factor two effect for the faint red galaxies, that could potentially be detected in observations.
  
\item Care should be given to consistently defining the environment for the original and shuffled samples. Utilizing separately the shuffled galaxies' environment can introduce systematics, as discussed in Appendix~\ref{sec:contam}. Even as such, our strong signal for faint red galaxies is only minimally affected, producing a robust signature of assembly bias.   

\end{itemize}

Compared with previous studies of the LF as a function of environment \citep{2005MNRAS.356.1155C,10.1093/mnras/stu1886, 2018MNRAS.476..741D}  and standard measurements of GAB via galaxy clustering (e.g., \citealt{2007MNRAS.374.1303C,2018ApJ...853...84Z}), our work not only reveals another facet of the environmental dependence of the LF, but also introduces a novel measure of assembly bias, which employs a fundamental statistic and yields strong signals. We view our approach as complementary to galaxy clustering studies of assembly bias, and plan to develop and explore practical applications in future work. We hope that this new perspective will inspire further investigations and motivate efforts to detect it in observations.

\section*{Data Availability}
The IllustrisTNG simulations, including TNG300, are publicly available and accessible at \url{www.tng-project.org/data} \citep{2019ComAC...6....2N}. Data directly related to this publication is available upon reasonable request.

\begin{acknowledgements}
I.Z. and S.C.(Contreras) acknowledge the hospitality of the ICC in Durham, where this work was initiated.  We thank Samuel Moore and Kai Wang for useful discussions. We appreciate the insightful comments and thoughtful suggestions provided by the anonymous referee, which improved the presentation of the paper. I.Z. was partially supported by a CWRU ACES+ Opportunity Grant. S.C.(Contreras) acknowledges the support of the ``Juan de la Cierva Incorporation'' fellowship (IJC2020-045705-I). S.C.(Cole) and P.N. acknowledge the support of STFC consolidated grant (ST/X001075/1).  
\end{acknowledgements}

\vspace{1cm}

\bibliography{LFpaper}{}

\begin{thebibliography}{}
\expandafter\ifx\csname natexlab\endcsname\relax\def\natexlab#1{#1}\fi
\providecommand{\url}[1]{\href{#1}{#1}}
\providecommand{\dodoi}[1]{doi:~\href{http://doi.org/#1}{\nolinkurl{#1}}}
\providecommand{\doeprint}[1]{\href{http://ascl.net/#1}{\nolinkurl{http://ascl.net/#1}}}
\providecommand{\doarXiv}[1]{\href{https://arxiv.org/abs/#1}{\nolinkurl{https://arxiv.org/abs/#1}}}

\bibitem[{{Alonso} {et~al.}(2015){Alonso}, {Eardley}, \& {Peacock}}]{2015MNRAS.447.2683A}
{Alonso}, D., {Eardley}, E., \& {Peacock}, J.~A. 2015, \mnras, 447, 2683, \dodoi{10.1093/mnras/stu2632}

\bibitem[{{Alpaslan} \& {Tinker}(2020)}]{2020MNRAS.496.5463A}
{Alpaslan}, M., \& {Tinker}, J.~L. 2020, \mnras, 496, 5463, \dodoi{10.1093/mnras/staa1844}

\bibitem[{{Artale} {et~al.}(2018){Artale}, {Zehavi}, {Contreras}, \& {Norberg}}]{2018MNRAS.480.3978A}
{Artale}, M.~C., {Zehavi}, I., {Contreras}, S., \& {Norberg}, P. 2018, \mnras, 480, 3978, \dodoi{10.1093/mnras/sty2110}

\bibitem[{{Benson} {et~al.}(2003){Benson}, {Bower}, {Frenk}, {Lacey}, {Baugh}, \& {Cole}}]{2003ApJ...599...38B}
{Benson}, A.~J., {Bower}, R.~G., {Frenk}, C.~S., {et~al.} 2003, \apj, 599, 38, \dodoi{10.1086/379160}

\bibitem[{{Blanton} \& {Berlind}(2007)}]{2007ApJ...664..791B}
{Blanton}, M.~R., \& {Berlind}, A.~A. 2007, \apj, 664, 791, \dodoi{10.1086/512478}

\bibitem[{{Blanton} {et~al.}(2005){Blanton}, {Eisenstein}, {Hogg}, {Schlegel}, \& {Brinkmann}}]{2005ApJ...629..143B}
{Blanton}, M.~R., {Eisenstein}, D., {Hogg}, D.~W., {Schlegel}, D.~J., \& {Brinkmann}, J. 2005, \apj, 629, 143, \dodoi{10.1086/422897}

\bibitem[{{Blanton} {et~al.}(2003){Blanton}, {Hogg}, {Bahcall}, {Brinkmann}, {Britton}, {Connolly}, {Csabai}, {Fukugita}, {Loveday}, {Meiksin}, {Munn}, {Nichol}, {Okamura}, {Quinn}, {Schneider}, {Shimasaku}, {Strauss}, {Tegmark}, {Vogeley}, \& {Weinberg}}]{2003ApJ...592..819B}
{Blanton}, M.~R., {Hogg}, D.~W., {Bahcall}, N.~A., {et~al.} 2003, \apj, 592, 819, \dodoi{10.1086/375776}

\bibitem[{{Bond} {et~al.}(1991){Bond}, {Cole}, {Efstathiou}, \& {Kaiser}}]{1991ApJ...379..440B}
{Bond}, J.~R., {Cole}, S., {Efstathiou}, G., \& {Kaiser}, N. 1991, \apj, 379, 440, \dodoi{10.1086/170520}

\bibitem[{{Bose} {et~al.}(2019){Bose}, {Eisenstein}, {Hernquist}, {Pillepich}, {Nelson}, {Marinacci}, {Springel}, \& {Vogelsberger}}]{2019MNRAS.490.5693B}
{Bose}, S., {Eisenstein}, D.~J., {Hernquist}, L., {et~al.} 2019, \mnras, 490, 5693, \dodoi{10.1093/mnras/stz2546}

\bibitem[{Burchett {et~al.}(2020)Burchett, Elek, Tejos, \& et~al.}]{Burchett2020}
Burchett, J.~N., Elek, O., Tejos, N., \& et~al. 2020, The Astrophysical Journal Letters, 891, L35, \dodoi{10.3847/2041-8213/ab700c}

\bibitem[{{Campbell} {et~al.}(2015){Campbell}, {van den Bosch}, {Hearin}, {Padmanabhan}, {Berlind}, {Mo}, {Tinker}, \& {Yang}}]{2015MNRAS.452..444C}
{Campbell}, D., {van den Bosch}, F.~C., {Hearin}, A., {et~al.} 2015, \mnras, 452, 444, \dodoi{10.1093/mnras/stv1091}

\bibitem[{{Chaves-Montero} {et~al.}(2016){Chaves-Montero}, {Angulo}, {Schaye}, {Schaller}, {Crain}, {Furlong}, \& {Theuns}}]{2016MNRAS.460.3100C}
{Chaves-Montero}, J., {Angulo}, R.~E., {Schaye}, J., {et~al.} 2016, \mnras, 460, 3100, \dodoi{10.1093/mnras/stw1225}

\bibitem[{{Contreras} {et~al.}(2021){Contreras}, {Angulo}, \& {Zennaro}}]{2021MNRAS.504.5205C}
{Contreras}, S., {Angulo}, R.~E., \& {Zennaro}, M. 2021, \mnras, 504, 5205, \dodoi{10.1093/mnras/stab1170}

\bibitem[{{Contreras} {et~al.}(2023){Contreras}, {Chaves-Montero}, \& {Angulo}}]{2023MNRAS.525.3149C}
{Contreras}, S., {Chaves-Montero}, J., \& {Angulo}, R.~E. 2023, \mnras, 525, 3149, \dodoi{10.1093/mnras/stad2434}

\bibitem[{{Contreras} {et~al.}(2019){Contreras}, {Zehavi}, {Padilla}, {Baugh}, {Jim{\'e}nez}, \& {Lacerna}}]{2019MNRAS.484.1133C}
{Contreras}, S., {Zehavi}, I., {Padilla}, N., {et~al.} 2019, \mnras, 484, 1133, \dodoi{10.1093/mnras/stz018}

\bibitem[{{Croton} \& {Farrar}(2008)}]{2008MNRAS.386.2285C}
{Croton}, D.~J., \& {Farrar}, G.~R. 2008, \mnras, 386, 2285, \dodoi{10.1111/j.1365-2966.2008.13204.x}

\bibitem[{{Croton} {et~al.}(2007){Croton}, {Gao}, \& {White}}]{2007MNRAS.374.1303C}
{Croton}, D.~J., {Gao}, L., \& {White}, S. D.~M. 2007, \mnras, 374, 1303, \dodoi{10.1111/j.1365-2966.2006.11230.x}

\bibitem[{{Croton} {et~al.}(2005){Croton}, {Farrar}, {Norberg}, {Colless}, {Peacock}, {Baldry}, {Baugh}, {Bland-Hawthorn}, {Bridges}, {Cannon}, {Cole}, {Collins}, {Couch}, {Dalton}, {De Propris}, {Driver}, {Efstathiou}, {Ellis}, {Frenk}, {Glazebrook}, {Jackson}, {Lahav}, {Lewis}, {Lumsden}, {Maddox}, {Madgwick}, {Peterson}, {Sutherland}, \& {Taylor}}]{2005MNRAS.356.1155C}
{Croton}, D.~J., {Farrar}, G.~R., {Norberg}, P., {et~al.} 2005, \mnras, 356, 1155, \dodoi{10.1111/j.1365-2966.2004.08546.x}

\bibitem[{{Davis} {et~al.}(1985){Davis}, {Efstathiou}, {Frenk}, \& {White}}]{1985ApJ...292..371D}
{Davis}, M., {Efstathiou}, G., {Frenk}, C.~S., \& {White}, S.~D.~M. 1985, \apj, 292, 371, \dodoi{10.1086/163168}

\bibitem[{{De Propris} {et~al.}(2013){De Propris}, {Phillipps}, \& {Bremer}}]{2013MNRAS.434.3469D}
{De Propris}, R., {Phillipps}, S., \& {Bremer}, M.~N. 2013, \mnras, 434, 3469, \dodoi{10.1093/mnras/stt1262}

\bibitem[{De~Propris {et~al.}(2003)De~Propris, Colless, Driver, Couch, Peacock, Baldry, Baugh, Bland-Hawthorn, Bridges, Cannon, Cole, Collins, Cross, Dalton, Efstathiou, Ellis, Frenk, Glazebrook, Hawkins, Jackson, Lahav, Lewis, Lumsden, Maddox, Madgwick, Norberg, Percival, Peterson, Sutherland, \& Taylor}]{10.1046/j.1365-8711.2003.06510.x}
De~Propris, R., Colless, M., Driver, S.~P., {et~al.} 2003, \mnras, 342, 725, \dodoi{10.1046/j.1365-8711.2003.06510.x}

\bibitem[{{Dragomir} {et~al.}(2018){Dragomir}, {Rodr{\'\i}guez-Puebla}, {Primack}, \& {Lee}}]{2018MNRAS.476..741D}
{Dragomir}, R., {Rodr{\'\i}guez-Puebla}, A., {Primack}, J.~R., \& {Lee}, C.~T. 2018, \mnras, 476, 741, \dodoi{10.1093/mnras/sty283}

\bibitem[{{Dressler}(1980)}]{1980ApJ...236..351D}
{Dressler}, A. 1980, \apj, 236, 351, \dodoi{10.1086/157753}

\bibitem[{{Eardley} {et~al.}(2015){Eardley}, {Peacock}, {McNaught-Roberts}, {Heymans}, {Norberg}, {Alpaslan}, {Baldry}, {Bland-Hawthorn}, {Brough}, {Cluver}, {Driver}, {Farrow}, {Liske}, {Loveday}, \& {Robotham}}]{2015MNRAS.448.3665E}
{Eardley}, E., {Peacock}, J.~A., {McNaught-Roberts}, T., {et~al.} 2015, \mnras, 448, 3665, \dodoi{10.1093/mnras/stv237}

\bibitem[{{Efstathiou} {et~al.}(1988){Efstathiou}, {Ellis}, \& {Peterson}}]{1988MNRAS.232..431E}
{Efstathiou}, G., {Ellis}, R.~S., \& {Peterson}, B.~A. 1988, \mnras, 232, 431, \dodoi{10.1093/mnras/232.2.431}

\bibitem[{{Gao} {et~al.}(2005){Gao}, {Springel}, \& {White}}]{2005MNRAS.363L..66G}
{Gao}, L., {Springel}, V., \& {White}, S. D.~M. 2005, \mnras, 363, L66, \dodoi{10.1111/j.1745-3933.2005.00084.x}

\bibitem[{Gao \& White(2007)}]{10.1111/j.1745-3933.2007.00292.x}
Gao, L., \& White, S. D.~M. 2007, \mnras, 377, L5, \dodoi{10.1111/j.1745-3933.2007.00292.x}

\bibitem[{{Genel} {et~al.}(2014){Genel}, {Vogelsberger}, {Springel}, {Sijacki}, {Nelson}, {Snyder}, {Rodriguez-Gomez}, {Torrey}, \& {Hernquist}}]{2014MNRAS.445..175G}
{Genel}, S., {Vogelsberger}, M., {Springel}, V., {et~al.} 2014, \mnras, 445, 175, \dodoi{10.1093/mnras/stu1654}

\bibitem[{{G{\'o}mez} {et~al.}(2003){G{\'o}mez}, {Nichol}, {Miller}, {Balogh}, {Goto}, {Zabludoff}, {Romer}, {Bernardi}, {Sheth}, {Hopkins}, {Castander}, {Connolly}, {Schneider}, {Brinkmann}, {Lamb}, {SubbaRao}, \& {York}}]{2003ApJ...584..210G}
{G{\'o}mez}, P.~L., {Nichol}, R.~C., {Miller}, C.~J., {et~al.} 2003, \apj, 584, 210, \dodoi{10.1086/345593}

\bibitem[{{Hadzhiyska} {et~al.}(2020){Hadzhiyska}, {Bose}, {Eisenstein}, {Hernquist}, \& {Spergel}}]{2020MNRAS.493.5506H}
{Hadzhiyska}, B., {Bose}, S., {Eisenstein}, D., {Hernquist}, L., \& {Spergel}, D.~N. 2020, \mnras, 493, 5506, \dodoi{10.1093/mnras/staa623}

\bibitem[{{Hadzhiyska} {et~al.}(2023){Hadzhiyska}, {Eisenstein}, {Hernquist}, {Pakmor}, {Bose}, {Delgado}, {Contreras}, {Kannan}, {White}, {Springel}, {Frenk}, {Hern{\'a}ndez-Aguayo}, {Barrera}, \& {Monica}}]{2023MNRAS.524.2507H}
{Hadzhiyska}, B., {Eisenstein}, D., {Hernquist}, L., {et~al.} 2023, \mnras, 524, 2507, \dodoi{10.1093/mnras/stad731}

\bibitem[{{Han} {et~al.}(2019){Han}, {Li}, {Jing}, {Nishimichi}, {Wang}, \& {Jiang}}]{2019MNRAS.482.1900H}
{Han}, J., {Li}, Y., {Jing}, Y., {et~al.} 2019, \mnras, 482, 1900, \dodoi{10.1093/mnras/sty2822}

\bibitem[{{Hearin} {et~al.}(2015){Hearin}, {Watson}, \& {van den Bosch}}]{2015MNRAS.452.1958H}
{Hearin}, A.~P., {Watson}, D.~F., \& {van den Bosch}, F.~C. 2015, \mnras, 452, 1958, \dodoi{10.1093/mnras/stv1358}

\bibitem[{{Hogg} {et~al.}(2003){Hogg}, {Blanton}, {Eisenstein}, {Gunn}, {Schlegel}, {Zehavi}, {Bahcall}, {Brinkmann}, {Csabai}, {Schneider}, {Weinberg}, \& {York}}]{2003ApJ...585L...5H}
{Hogg}, D.~W., {Blanton}, M.~R., {Eisenstein}, D.~J., {et~al.} 2003, \apjl, 585, L5, \dodoi{10.1086/374238}

\bibitem[{Hoyle {et~al.}(2005)Hoyle, Rojas, Vogeley, \& Brinkmann}]{Hoyle_2005}
Hoyle, F., Rojas, R.~R., Vogeley, M.~S., \& Brinkmann, J. 2005, \apj, 620, 618–628, \dodoi{10.1086/427176}

\bibitem[{{Kakos} {et~al.}(2024){Kakos}, {Rodr{\'\i}guez-Puebla}, {Primack}, {Faber}, {Koo}, {Behroozi}, \& {Avila-Reese}}]{Kakos2024}
{Kakos}, J., {Rodr{\'\i}guez-Puebla}, A., {Primack}, J.~R., {et~al.} 2024, \mnras, 533, 3585, \dodoi{10.1093/mnras/stae1969}

\bibitem[{{Kauffmann} {et~al.}(2004){Kauffmann}, {White}, {Heckman}, {M{\'e}nard}, {Brinchmann}, {Charlot}, {Tremonti}, \& {Brinkmann}}]{2004MNRAS.353..713K}
{Kauffmann}, G., {White}, S. D.~M., {Heckman}, T.~M., {et~al.} 2004, \mnras, 353, 713, \dodoi{10.1111/j.1365-2966.2004.08117.x}

\bibitem[{{Kennicutt}(1983)}]{1983AJ.....88..483K}
{Kennicutt}, R.~C., J. 1983, \aj, 88, 483, \dodoi{10.1086/113334}

\bibitem[{{Lemson} \& {Kauffmann}(1999)}]{1999MNRAS.302..111L}
{Lemson}, G., \& {Kauffmann}, G. 1999, \mnras, 302, 111, \dodoi{10.1046/j.1365-8711.1999.02090.x}

\bibitem[{Lewis {et~al.}(2002)Lewis, Balogh, De~Propris, Couch, Bower, Offer, Bland-Hawthorn, Baldry, Baugh, Bridges, Cannon, Cole, Colless, Collins, Cross, Dalton, Driver, Efstathiou, Ellis, Frenk, Glazebrook, Hawkins, Jackson, Lahav, Lumsden, Maddox, Madgwick, Norberg, Peacock, Percival, Peterson, Sutherland, \& Taylor}]{10.1046/j.1365-8711.2002.05558.x}
Lewis, I., Balogh, M., De~Propris, R., {et~al.} 2002, \mnras, 334, 673, \dodoi{10.1046/j.1365-8711.2002.05558.x}

\bibitem[{{Lin} {et~al.}(1996){Lin}, {Yee}, {Carlberg}, \& {Ellingson}}]{1996JRASC..90..337L}
{Lin}, H., {Yee}, H.~K.~C., {Carlberg}, R.~G., \& {Ellingson}, E. 1996, \jrasc, 90, 337

\bibitem[{{Lin} {et~al.}(2016){Lin}, {Mandelbaum}, {Huang}, {Huang}, {Dalal}, {Diemer}, {Jian}, \& {Kravtsov}}]{2016ApJ...819..119L}
{Lin}, Y.-T., {Mandelbaum}, R., {Huang}, Y.-H., {et~al.} 2016, \apj, 819, 119, \dodoi{10.3847/0004-637X/819/2/119}

\bibitem[{{Loveday} {et~al.}(2012){Loveday}, {Norberg}, {Baldry}, {Driver}, {Hopkins}, {Peacock}, {Bamford}, {Liske}, {Bland-Hawthorn}, {Brough}, {Brown}, {Cameron}, {Conselice}, {Croom}, {Frenk}, {Gunawardhana}, {Hill}, {Jones}, {Kelvin}, {Kuijken}, {Nichol}, {Parkinson}, {Phillipps}, {Pimbblet}, {Popescu}, {Prescott}, {Robotham}, {Sharp}, {Sutherland}, {Taylor}, {Thomas}, {Tuffs}, {van Kampen}, \& {Wijesinghe}}]{2012MNRAS.420.1239L}
{Loveday}, J., {Norberg}, P., {Baldry}, I.~K., {et~al.} 2012, \mnras, 420, 1239, \dodoi{10.1111/j.1365-2966.2011.20111.x}

\bibitem[{Loveday {et~al.}(2012)Loveday, Norberg, Baldry, Driver, Hopkins, Peacock, Bamford, Liske, Bland-Hawthorn, Brough, Brown, Cameron, Conselice, Croom, Frenk, Gunawardhana, Hill, Jones, Kelvin, Kuijken, Nichol, Parkinson, Phillipps, Pimbblet, Popescu, Prescott, Robotham, Sharp, Sutherland, Taylor, Thomas, Tuffs, van Kampen, \& Wijesinghe}]{10.1111/j.1365-2966.2011.20111.x}
Loveday, J., Norberg, P., Baldry, I.~K., {et~al.} 2012, \mnras, 420, 1239, \dodoi{10.1111/j.1365-2966.2011.20111.x}

\bibitem[{{Lupton}(1993)}]{1993stp..book.....L}
{Lupton}, R. 1993, {Statistics in Theory and Practice} (Princeton: Princeton University Press)

\bibitem[{Mahajan \& Raychaudhury(2009)}]{10.1111/j.1365-2966.2009.15512.x}
Mahajan, S., \& Raychaudhury, S. 2009, \mnras, 400, 687, \dodoi{10.1111/j.1365-2966.2009.15512.x}

\bibitem[{Maller {et~al.}(2009)Maller, Berlind, Blanton, \& Hogg}]{Maller_2009}
Maller, A.~H., Berlind, A.~A., Blanton, M.~R., \& Hogg, D.~W. 2009, \apj, 691, 394–406, \dodoi{10.1088/0004-637x/691/1/394}

\bibitem[{{Marinacci} {et~al.}(2018){Marinacci}, {Vogelsberger}, {Pakmor}, {Torrey}, {Springel}, {Hernquist}, {Nelson}, {Weinberger}, {Pillepich}, {Naiman}, \& {Genel}}]{2018MNRAS.480.5113M}
{Marinacci}, F., {Vogelsberger}, M., {Pakmor}, R., {et~al.} 2018, \mnras, 480, 5113, \dodoi{10.1093/mnras/sty2206}

\bibitem[{McNaught-Roberts {et~al.}(2014)McNaught-Roberts, Norberg, Baugh, Lacey, Loveday, Peacock, Baldry, Bland-Hawthorn, Brough, Driver, Robotham, \& VÃ¡zquez-Mata}]{10.1093/mnras/stu1886}
McNaught-Roberts, T., Norberg, P., Baugh, C., {et~al.} 2014, \mnras, 445, 2125, \dodoi{10.1093/mnras/stu1886}

\bibitem[{{Mo} {et~al.}(2004){Mo}, {Yang}, {van den Bosch}, \& {Jing}}]{2004MNRAS.349..205M}
{Mo}, H.~J., {Yang}, X., {van den Bosch}, F.~C., \& {Jing}, Y.~P. 2004, \mnras, 349, 205, \dodoi{10.1111/j.1365-2966.2004.07485.x}

\bibitem[{{Montero-Dorta} {et~al.}(2025){Montero-Dorta}, {Contreras}, {Celeste Artale}, {Rodriguez}, \& {Favole}}]{2025A&A...695A.159M}
{Montero-Dorta}, A.~D., {Contreras}, S., {Celeste Artale}, M., {Rodriguez}, F., \& {Favole}, G. 2025, \aap, 695, A159, \dodoi{10.1051/0004-6361/202452709}

\bibitem[{{Muldrew} {et~al.}(2012){Muldrew}, {Croton}, {Skibba}, {Pearce}, {Ann}, {Baldry}, {Brough}, {Choi}, {Conselice}, {Cowan}, {Gallazzi}, {Gray}, {Gr{\"u}tzbauch}, {Li}, {Park}, {Pilipenko}, {Podgorzec}, {Robotham}, {Wilman}, {Yang}, {Zhang}, \& {Zibetti}}]{2012MNRAS.419.2670M}
{Muldrew}, S.~I., {Croton}, D.~J., {Skibba}, R.~A., {et~al.} 2012, \mnras, 419, 2670, \dodoi{10.1111/j.1365-2966.2011.19922.x}

\bibitem[{{Nelson} {et~al.}(2018){Nelson}, {Pillepich}, {Springel}, {Weinberger}, {Hernquist}, {Pakmor}, {Genel}, {Torrey}, {Vogelsberger}, {Kauffmann}, {Marinacci}, \& {Naiman}}]{2018MNRAS.475..624N}
{Nelson}, D., {Pillepich}, A., {Springel}, V., {et~al.} 2018, \mnras, 475, 624, \dodoi{10.1093/mnras/stx3040}

\bibitem[{{Nelson} {et~al.}(2019){Nelson}, {Springel}, {Pillepich}, {Rodriguez-Gomez}, {Torrey}, {Genel}, {Vogelsberger}, {Pakmor}, {Marinacci}, {Weinberger}, {Kelley}, {Lovell}, {Diemer}, \& {Hernquist}}]{2019ComAC...6....2N}
{Nelson}, D., {Springel}, V., {Pillepich}, A., {et~al.} 2019, Computational Astrophysics and Cosmology, 6, 2, \dodoi{10.1186/s40668-019-0028-x}

\bibitem[{{Norberg} {et~al.}(2009){Norberg}, {Baugh}, {Gazta{\~n}aga}, \& {Croton}}]{2009MNRAS.396...19N}
{Norberg}, P., {Baugh}, C.~M., {Gazta{\~n}aga}, E., \& {Croton}, D.~J. 2009, \mnras, 396, 19, \dodoi{10.1111/j.1365-2966.2009.14389.x}

\bibitem[{{Norberg} {et~al.}(2002){Norberg}, {Cole}, {Baugh}, {Frenk}, {Baldry}, {Bland-Hawthorn}, {Bridges}, {Cannon}, {Colless}, {Collins}, {Couch}, {Cross}, {Dalton}, {De Propris}, {Driver}, {Efstathiou}, {Ellis}, {Glazebrook}, {Jackson}, {Lahav}, {Lewis}, {Lumsden}, {Maddox}, {Madgwick}, {Peacock}, {Peterson}, {Sutherland}, {Taylor}, \& {2DFGRS Team}}]{2002MNRAS.336..907N}
{Norberg}, P., {Cole}, S., {Baugh}, C.~M., {et~al.} 2002, \mnras, 336, 907, \dodoi{10.1046/j.1365-8711.2002.05831.x}

\bibitem[{{Obuljen} {et~al.}(2020){Obuljen}, {Percival}, \& {Dalal}}]{2020JCAP...10..058O}
{Obuljen}, A., {Percival}, W.~J., \& {Dalal}, N. 2020, \jcap, 2020, 058, \dodoi{10.1088/1475-7516/2020/10/058}

\bibitem[{{Oemler}(1974)}]{1974ApJ...194....1O}
{Oemler}, Augustus, J. 1974, \apj, 194, 1, \dodoi{10.1086/153216}

\bibitem[{{Ortega-Martinez} {et~al.}(2024){Ortega-Martinez}, {Contreras}, {Angulo}, \& {Chaves-Montero}}]{2024arXiv241111830O}
{Ortega-Martinez}, S., {Contreras}, S., {Angulo}, R.~E., \& {Chaves-Montero}, J. 2024, arXiv e-prints, arXiv:2411.11830, \dodoi{10.48550/arXiv.2411.11830}

\bibitem[{{Oyarz{\'u}n} {et~al.}(2024){Oyarz{\'u}n}, {Tinker}, {Bundy}, {Xhakaj}, \& {Wyithe}}]{2024ApJ...974...29O}
{Oyarz{\'u}n}, G.~A., {Tinker}, J.~L., {Bundy}, K., {Xhakaj}, E., \& {Wyithe}, J. S.~B. 2024, \apj, 974, 29, \dodoi{10.3847/1538-4357/ad6de1}

\bibitem[{{Pearl} {et~al.}(2024){Pearl}, {Zentner}, {Newman}, {Bezanson}, {Wang}, {Moustakas}, {Aguilar}, {Ahlen}, {Brooks}, {Claybaugh}, {Cole}, {Dawson}, {de la Macorra}, {Doel}, {Forero-Romero}, {Gontcho A Gontcho}, {Honscheid}, {Landriau}, {Manera}, {Martini}, {Meisner}, {Miquel}, {Nie}, {Percival}, {Prada}, {Rezaie}, {Rossi}, {Sanchez}, {Schubnell}, {Tarl{\'e}}, {Weaver}, \& {Zhou}}]{2024ApJ...963..116P}
{Pearl}, A.~N., {Zentner}, A.~R., {Newman}, J.~A., {et~al.} 2024, \apj, 963, 116, \dodoi{10.3847/1538-4357/ad1ffd}

\bibitem[{{Peng} {et~al.}(2010){Peng}, {Lilly}, {Kova{\v{c}}}, {Bolzonella}, {Pozzetti}, {Renzini}, {Zamorani}, {Ilbert}, {Knobel}, {Iovino}, {Maier}, {Cucciati}, {Tasca}, {Carollo}, {Silverman}, {Kampczyk}, {de Ravel}, {Sanders}, {Scoville}, {Contini}, {Mainieri}, {Scodeggio}, {Kneib}, {Le F{\`e}vre}, {Bardelli}, {Bongiorno}, {Caputi}, {Coppa}, {de la Torre}, {Franzetti}, {Garilli}, {Lamareille}, {Le Borgne}, {Le Brun}, {Mignoli}, {Perez Montero}, {Pello}, {Ricciardelli}, {Tanaka}, {Tresse}, {Vergani}, {Welikala}, {Zucca}, {Oesch}, {Abbas}, {Barnes}, {Bordoloi}, {Bottini}, {Cappi}, {Cassata}, {Cimatti}, {Fumana}, {Hasinger}, {Koekemoer}, {Leauthaud}, {Maccagni}, {Marinoni}, {McCracken}, {Memeo}, {Meneux}, {Nair}, {Porciani}, {Presotto}, \& {Scaramella}}]{2010ApJ...721..193P}
{Peng}, Y.-j., {Lilly}, S.~J., {Kova{\v{c}}}, K., {et~al.} 2010, \apj, 721, 193, \dodoi{10.1088/0004-637X/721/1/193}

\bibitem[{Pillepich {et~al.}(2017)Pillepich, Nelson, Hernquist, Springel, Pakmor, Torrey, Weinberger, Genel, Naiman, Marinacci, \& Vogelsberger}]{10.1093/mnras/stx3112}
Pillepich, A., Nelson, D., Hernquist, L., {et~al.} 2017, \mnras, 475, 648, \dodoi{10.1093/mnras/stx3112}

\bibitem[{{Planck Collaboration} {et~al.}(2016){Planck Collaboration}, {Ade}, {Aghanim}, {Arnaud}, {Ashdown}, {Aumont}, {Baccigalupi}, {Banday}, {Barreiro}, {Bartlett}, {Bartolo}, {Battaner}, {Battye}, {Benabed}, {Beno{\^\i}t}, {Benoit-L{\'e}vy}, {Bernard}, {Bersanelli}, {Bielewicz}, {Bock}, {Bonaldi}, {Bonavera}, {Bond}, {Borrill}, {Bouchet}, {Boulanger}, {Bucher}, {Burigana}, {Butler}, {Calabrese}, {Cardoso}, {Catalano}, {Challinor}, {Chamballu}, {Chary}, {Chiang}, {Chluba}, {Christensen}, {Church}, {Clements}, {Colombi}, {Colombo}, {Combet}, {Coulais}, {Crill}, {Curto}, {Cuttaia}, {Danese}, {Davies}, {Davis}, {de Bernardis}, {de Rosa}, {de Zotti}, {Delabrouille}, {D{\'e}sert}, {Di Valentino}, {Dickinson}, {Diego}, {Dolag}, {Dole}, {Donzelli}, {Dor{\'e}}, {Douspis}, {Ducout}, {Dunkley}, {Dupac}, {Efstathiou}, {Elsner}, {En{\ss}lin}, {Eriksen}, {Farhang}, {Fergusson}, {Finelli}, {Forni}, {Frailis}, {Fraisse}, {Franceschi}, {Frejsel}, {Galeotta}, {Galli}, {Ganga}, {Gauthier}, {Gerbino}, {Ghosh}, {Giard},
  {Giraud-H{\'e}raud}, {Giusarma}, {Gjerl{\o}w}, {Gonz{\'a}lez-Nuevo}, {G{\'o}rski}, {Gratton}, {Gregorio}, {Gruppuso}, {Gudmundsson}, {Hamann}, {Hansen}, {Hanson}, {Harrison}, {Helou}, {Henrot-Versill{\'e}}, {Hern{\'a}ndez-Monteagudo}, {Herranz}, {Hildebrandt}, {Hivon}, {Hobson}, {Holmes}, {Hornstrup}, {Hovest}, {Huang}, {Huffenberger}, {Hurier}, {Jaffe}, {Jaffe}, {Jones}, {Juvela}, {Keih{\"a}nen}, {Keskitalo}, {Kisner}, {Kneissl}, {Knoche}, {Knox}, {Kunz}, {Kurki-Suonio}, {Lagache}, {L{\"a}hteenm{\"a}ki}, {Lamarre}, {Lasenby}, {Lattanzi}, {Lawrence}, {Leahy}, {Leonardi}, {Lesgourgues}, {Levrier}, {Lewis}, {Liguori}, {Lilje}, {Linden-V{\o}rnle}, {L{\'o}pez-Caniego}, {Lubin}, {Mac{\'\i}as-P{\'e}rez}, {Maggio}, {Maino}, {Mandolesi}, {Mangilli}, {Marchini}, {Maris}, {Martin}, {Martinelli}, {Mart{\'\i}nez-Gonz{\'a}lez}, {Masi}, {Matarrese}, {McGehee}, {Meinhold}, {Melchiorri}, {Melin}, {Mendes}, {Mennella}, {Migliaccio}, {Millea}, {Mitra}, {Miville-Desch{\^e}nes}, {Moneti}, {Montier}, {Morgante}, {Mortlock},
  {Moss}, {Munshi}, {Murphy}, {Naselsky}, {Nati}, {Natoli}, {Netterfield}, {N{\o}rgaard-Nielsen}, {Noviello}, {Novikov}, {Novikov}, {Oxborrow}, {Paci}, {Pagano}, {Pajot}, {Paladini}, {Paoletti}, {Partridge}, {Pasian}, {Patanchon}, {Pearson}, {Perdereau}, {Perotto}, {Perrotta}, {Pettorino}, {Piacentini}, {Piat}, {Pierpaoli}, {Pietrobon}, {Plaszczynski}, {Pointecouteau}, {Polenta}, {Popa}, {Pratt}, {Pr{\'e}zeau}, {Prunet}, {Puget}, {Rachen}, {Reach}, {Rebolo}, {Reinecke}, {Remazeilles}, {Renault}, {Renzi}, {Ristorcelli}, {Rocha}, {Rosset}, {Rossetti}, {Roudier}, {Rouill{\'e} d'Orfeuil}, {Rowan-Robinson}, {Rubi{\~n}o-Mart{\'\i}n}, {Rusholme}, {Said}, {Salvatelli}, {Salvati}, {Sandri}, {Santos}, {Savelainen}, {Savini}, {Scott}, {Seiffert}, {Serra}, {Shellard}, {Spencer}, {Spinelli}, {Stolyarov}, {Stompor}, {Sudiwala}, {Sunyaev}, {Sutton}, {Suur-Uski}, {Sygnet}, {Tauber}, {Terenzi}, {Toffolatti}, {Tomasi}, {Tristram}, {Trombetti}, {Tucci}, {Tuovinen}, {T{\"u}rler}, {Umana}, {Valenziano}, {Valiviita}, {Van Tent},
  {Vielva}, {Villa}, {Wade}, {Wandelt}, {Wehus}, {White}, {White}, {Wilkinson}, {Yvon}, {Zacchei}, \& {Zonca}}]{2016A&A...594A..13P}
{Planck Collaboration}, {Ade}, P.~A.~R., {Aghanim}, N., {et~al.} 2016, \aap, 594, A13, \dodoi{10.1051/0004-6361/201525830}

\bibitem[{{Ramakrishnan} {et~al.}(2019){Ramakrishnan}, {Paranjape}, {Hahn}, \& {Sheth}}]{2019MNRAS.489.2977R}
{Ramakrishnan}, S., {Paranjape}, A., {Hahn}, O., \& {Sheth}, R.~K. 2019, \mnras, 489, 2977, \dodoi{10.1093/mnras/stz2344}

\bibitem[{{Robotham} {et~al.}(2006){Robotham}, {Wallace}, {Phillipps}, \& {De Propris}}]{2006ApJ...652.1077R}
{Robotham}, A., {Wallace}, C., {Phillipps}, S., \& {De Propris}, R. 2006, \apj, 652, 1077, \dodoi{10.1086/508130}

\bibitem[{Salcedo {et~al.}(2018)Salcedo, Maller, Berlind, Sinha, McBride, Behroozi, Wechsler, \& Weinberg}]{10.1093/mnras/sty109}
Salcedo, A.~N., Maller, A.~H., Berlind, A.~A., {et~al.} 2018, \mnras, 475, 4411, \dodoi{10.1093/mnras/sty109}

\bibitem[{{Sato-Polito} {et~al.}(2019){Sato-Polito}, {Montero-Dorta}, {Abramo}, {Prada}, \& {Klypin}}]{2019MNRAS.487.1570S}
{Sato-Polito}, G., {Montero-Dorta}, A.~D., {Abramo}, L.~R., {Prada}, F., \& {Klypin}, A. 2019, \mnras, 487, 1570, \dodoi{10.1093/mnras/stz1338}

\bibitem[{Sheth \& Tormen(2004)}]{10.1111/j.1365-2966.2004.07733.x}
Sheth, R.~K., \& Tormen, G. 2004, \mnras, 350, 1385, \dodoi{10.1111/j.1365-2966.2004.07733.x}

\bibitem[{{Sijacki} {et~al.}(2015){Sijacki}, {Vogelsberger}, {Genel}, {Springel}, {Torrey}, {Snyder}, {Nelson}, \& {Hernquist}}]{2015MNRAS.452..575S}
{Sijacki}, D., {Vogelsberger}, M., {Genel}, S., {et~al.} 2015, \mnras, 452, 575, \dodoi{10.1093/mnras/stv1340}

\bibitem[{{Sin} {et~al.}(2017){Sin}, {Lilly}, \& {Henriques}}]{2017MNRAS.471.1192S}
{Sin}, L. P.~T., {Lilly}, S.~J., \& {Henriques}, B. M.~B. 2017, \mnras, 471, 1192, \dodoi{10.1093/mnras/stx1674}

\bibitem[{{Sinha} \& {Garrison}(2020)}]{2020MNRAS.491.3022S}
{Sinha}, M., \& {Garrison}, L.~H. 2020, \mnras, 491, 3022, \dodoi{10.1093/mnras/stz3157}

\bibitem[{Springel(2010)}]{10.1111/j.1365-2966.2009.15715.x}
Springel, V. 2010, \mnras, 401, 791, \dodoi{10.1111/j.1365-2966.2009.15715.x}

\bibitem[{{Springel} {et~al.}(2001){Springel}, {White}, {Tormen}, \& {Kauffmann}}]{2001MNRAS.328..726S}
{Springel}, V., {White}, S. D.~M., {Tormen}, G., \& {Kauffmann}, G. 2001, \mnras, 328, 726, \dodoi{10.1046/j.1365-8711.2001.04912.x}

\bibitem[{{Springel} {et~al.}(2018){Springel}, {Pakmor}, {Pillepich}, {Weinberger}, {Nelson}, {Hernquist}, {Vogelsberger}, {Genel}, {Torrey}, {Marinacci}, \& {Naiman}}]{2018MNRAS.475..676S}
{Springel}, V., {Pakmor}, R., {Pillepich}, A., {et~al.} 2018, \mnras, 475, 676, \dodoi{10.1093/mnras/stx3304}

\bibitem[{{Sunayama} \& {More}(2019)}]{2019MNRAS.490.4945S}
{Sunayama}, T., \& {More}, S. 2019, \mnras, 490, 4945, \dodoi{10.1093/mnras/stz2832}

\bibitem[{{Vogelsberger} {et~al.}(2014){Vogelsberger}, {Genel}, {Springel}, {Torrey}, {Sijacki}, {Xu}, {Snyder}, {Nelson}, \& {Hernquist}}]{2014MNRAS.444.1518V}
{Vogelsberger}, M., {Genel}, S., {Springel}, V., {et~al.} 2014, \mnras, 444, 1518, \dodoi{10.1093/mnras/stu1536}

\bibitem[{{Walsh} \& {Tinker}(2019)}]{2019MNRAS.488..470W}
{Walsh}, K., \& {Tinker}, J. 2019, \mnras, 488, 470, \dodoi{10.1093/mnras/stz1351}

\bibitem[{{Wang} {et~al.}(2013){Wang}, {Weinmann}, {De Lucia}, \& {Yang}}]{2013MNRAS.433..515W}
{Wang}, L., {Weinmann}, S.~M., {De Lucia}, G., \& {Yang}, X. 2013, \mnras, 433, 515, \dodoi{10.1093/mnras/stt743}

\bibitem[{{Wechsler} \& {Tinker}(2018)}]{2018ARA&A..56..435W}
{Wechsler}, R.~H., \& {Tinker}, J.~L. 2018, \araa, 56, 435, \dodoi{10.1146/annurev-astro-081817-051756}

\bibitem[{{Wechsler} {et~al.}(2006){Wechsler}, {Zentner}, {Bullock}, {Kravtsov}, \& {Allgood}}]{2006ApJ...652...71W}
{Wechsler}, R.~H., {Zentner}, A.~R., {Bullock}, J.~S., {Kravtsov}, A.~V., \& {Allgood}, B. 2006, \apj, 652, 71, \dodoi{10.1086/507120}

\bibitem[{Wetzel {et~al.}(2012)Wetzel, Tinker, \& Conroy}]{10.1111/j.1365-2966.2012.21188.x}
Wetzel, A.~R., Tinker, J.~L., \& Conroy, C. 2012, \mnras, 424, 232, \dodoi{10.1111/j.1365-2966.2012.21188.x}

\bibitem[{{White}(1996)}]{1996clss.conf..349W}
{White}, S.~D.~M. 1996, in Cosmology and Large Scale Structure, ed. R.~{Schaeffer}, J.~{Silk}, M.~{Spiro}, \& J.~{Zinn-Justin}, 349

\bibitem[{{White}(1999)}]{1999Ap&SS.267..355W}
{White}, S.~D.~M. 1999, \apss, 267, 355, \dodoi{10.1023/A:1002770429758}

\bibitem[{{White} \& {Frenk}(1991)}]{1991ApJ...379...52W}
{White}, S. D.~M., \& {Frenk}, C.~S. 1991, \apj, 379, 52, \dodoi{10.1086/170483}

\bibitem[{{Xu} {et~al.}(2021{\natexlab{a}}){Xu}, {Kumar}, {Zehavi}, \& {Contreras}}]{2021MNRAS.507.4879X}
{Xu}, X., {Kumar}, S., {Zehavi}, I., \& {Contreras}, S. 2021{\natexlab{a}}, \mnras, 507, 4879, \dodoi{10.1093/mnras/stab2464}

\bibitem[{{Xu} {et~al.}(2021{\natexlab{b}}){Xu}, {Zehavi}, \& {Contreras}}]{2021MNRAS.502.3242X}
{Xu}, X., {Zehavi}, I., \& {Contreras}, S. 2021{\natexlab{b}}, \mnras, 502, 3242, \dodoi{10.1093/mnras/stab100}

\bibitem[{{Yang} {et~al.}(2003){Yang}, {Mo}, \& {van den Bosch}}]{2003MNRAS.339.1057Y}
{Yang}, X., {Mo}, H.~J., \& {van den Bosch}, F.~C. 2003, \mnras, 339, 1057, \dodoi{10.1046/j.1365-8711.2003.06254.x}

\bibitem[{{Yuan} {et~al.}(2021){Yuan}, {Hadzhiyska}, {Bose}, {Eisenstein}, \& {Guo}}]{2021MNRAS.502.3582Y}
{Yuan}, S., {Hadzhiyska}, B., {Bose}, S., {Eisenstein}, D.~J., \& {Guo}, H. 2021, \mnras, 502, 3582, \dodoi{10.1093/mnras/stab235}

\bibitem[{{Zehavi} {et~al.}(2018){Zehavi}, {Contreras}, {Padilla}, {Smith}, {Baugh}, \& {Norberg}}]{2018ApJ...853...84Z}
{Zehavi}, I., {Contreras}, S., {Padilla}, N., {et~al.} 2018, \apj, 853, 84, \dodoi{10.3847/1538-4357/aaa54a}

\bibitem[{{Zehavi} {et~al.}(2019){Zehavi}, {Kerby}, {Contreras}, {Jim{\'e}nez}, {Padilla}, \& {Baugh}}]{2019ApJ...887...17Z}
{Zehavi}, I., {Kerby}, S.~E., {Contreras}, S., {et~al.} 2019, \apj, 887, 17, \dodoi{10.3847/1538-4357/ab4d4d}

\bibitem[{{Zehavi} {et~al.}(2002){Zehavi}, {Blanton}, {Frieman}, {Weinberg}, {Mo}, {Strauss}, {Anderson}, {Annis}, {Bahcall}, {Bernardi}, {Briggs}, {Brinkmann}, {Burles}, {Carey}, {Castander}, {Connolly}, {Csabai}, {Dalcanton}, {Dodelson}, {Doi}, {Eisenstein}, {Evans}, {Finkbeiner}, {Friedman}, {Fukugita}, {Gunn}, {Hennessy}, {Hindsley}, {Ivezi{\'c}}, {Kent}, {Knapp}, {Kron}, {Kunszt}, {Lamb}, {Leger}, {Long}, {Loveday}, {Lupton}, {McKay}, {Meiksin}, {Merrelli}, {Munn}, {Narayanan}, {Newcomb}, {Nichol}, {Owen}, {Peoples}, {Pope}, {Rockosi}, {Schlegel}, {Schneider}, {Scoccimarro}, {Sheth}, {Siegmund}, {Smee}, {Snir}, {Stebbins}, {Stoughton}, {SubbaRao}, {Szalay}, {Szapudi}, {Tegmark}, {Tucker}, {Uomoto}, {Vanden Berk}, {Vogeley}, {Waddell}, {Yanny}, \& {York}}]{2002ApJ...571..172Z}
{Zehavi}, I., {Blanton}, M.~R., {Frieman}, J.~A., {et~al.} 2002, \apj, 571, 172, \dodoi{10.1086/339893}

\bibitem[{{Zentner} {et~al.}(2014){Zentner}, {Hearin}, \& {van den Bosch}}]{2014MNRAS.443.3044Z}
{Zentner}, A.~R., {Hearin}, A.~P., \& {van den Bosch}, F.~C. 2014, \mnras, 443, 3044, \dodoi{10.1093/mnras/stu1383}

\bibitem[{{Zhu} {et~al.}(2006){Zhu}, {Zheng}, {Lin}, {Jing}, {Kang}, \& {Gao}}]{2006ApJ...639L...5Z}
{Zhu}, G., {Zheng}, Z., {Lin}, W.~P., {et~al.} 2006, \apjl, 639, L5, \dodoi{10.1086/501501}

\bibitem[{{Zu} \& {Mandelbaum}(2016)}]{2016MNRAS.457.4360Z}
{Zu}, Y., \& {Mandelbaum}, R. 2016, \mnras, 457, 4360, \dodoi{10.1093/mnras/stw221}

\bibitem[{{Zu} {et~al.}(2017){Zu}, {Mandelbaum}, {Simet}, {Rozo}, \& {Rykoff}}]{2017MNRAS.470..551Z}
{Zu}, Y., {Mandelbaum}, R., {Simet}, M., {Rozo}, E., \& {Rykoff}, E.~S. 2017, \mnras, 470, 551, \dodoi{10.1093/mnras/stx1264}

\bibitem[{{Zu} {et~al.}(2008){Zu}, {Zheng}, {Zhu}, \& {Jing}}]{2008ApJ...686...41Z}
{Zu}, Y., {Zheng}, Z., {Zhu}, G., \& {Jing}, Y.~P. 2008, \apj, 686, 41, \dodoi{10.1086/591071}

\end{thebibliography}
\bibliographystyle{aasjournal}

\appendix

\section{Impact of using the Environment of Shuffled Samples}
\label{sec:contam}

As described in \S~\ref{sec:methods}, we measure the environment using the galaxy positions, as a practical proxy for the underlying dark matter field. The galaxy density $\delta_8$ can be readily obtained from observations, and used to define the large-scale environment of the halos and identify the different environment regimes used in our analysis.  The shuffling mechanism, while changing the positions of galaxies within halos of fixed mass, does not change the underlying dark matter distribution or other halo properties.  Thus, in our analysis presented in the main text, we use the same $\delta_8$ measures associated with the halos throughout.  However, shuffling the galaxy sample does, in practice, slightly alter the environment (since assembly bias effects are tied to environment). So one might be concerned how this impacts our measurements. We stress that our theoretical predictions for assembly bias in the environment-dependent LF are not impacted by this (in the same way that it would be immaterial had we calculated the density from the dark matter or halos distribution). We associate the halos to the five environment bins and examine the contribution of assembly bias to the LF measured in each bin, so the halos' association to the density bins should not be changed in the process.  However, the change of environment with the shuffling may be relevant to future applications to model and detect these effects in real data, so it is worth considering here.

Measuring the environment from a shuffled spatial distribution of galaxies results in modified environment estimates, with subtle but possibly important changes in the values of $\delta_8$. We refer to this new measured environment as the `shuffled environment', in contrast with the `original environment' measured from the original TNG300 galaxy sample. In this context, the standard measurement uses the original environment for both the original the shuffled samples. As mentioned in \S~\ref{sec:result1}, we have also confirmed that if we switch to using a shuffled environment consistently for {\it both} samples, the results remain very similar. In what follows, we investigate what happens if we use the original environment for the original sample but use the shuffled environment for the shuffled sample.

\begin{figure}
\centering 
\includegraphics[width=0.49\columnwidth]{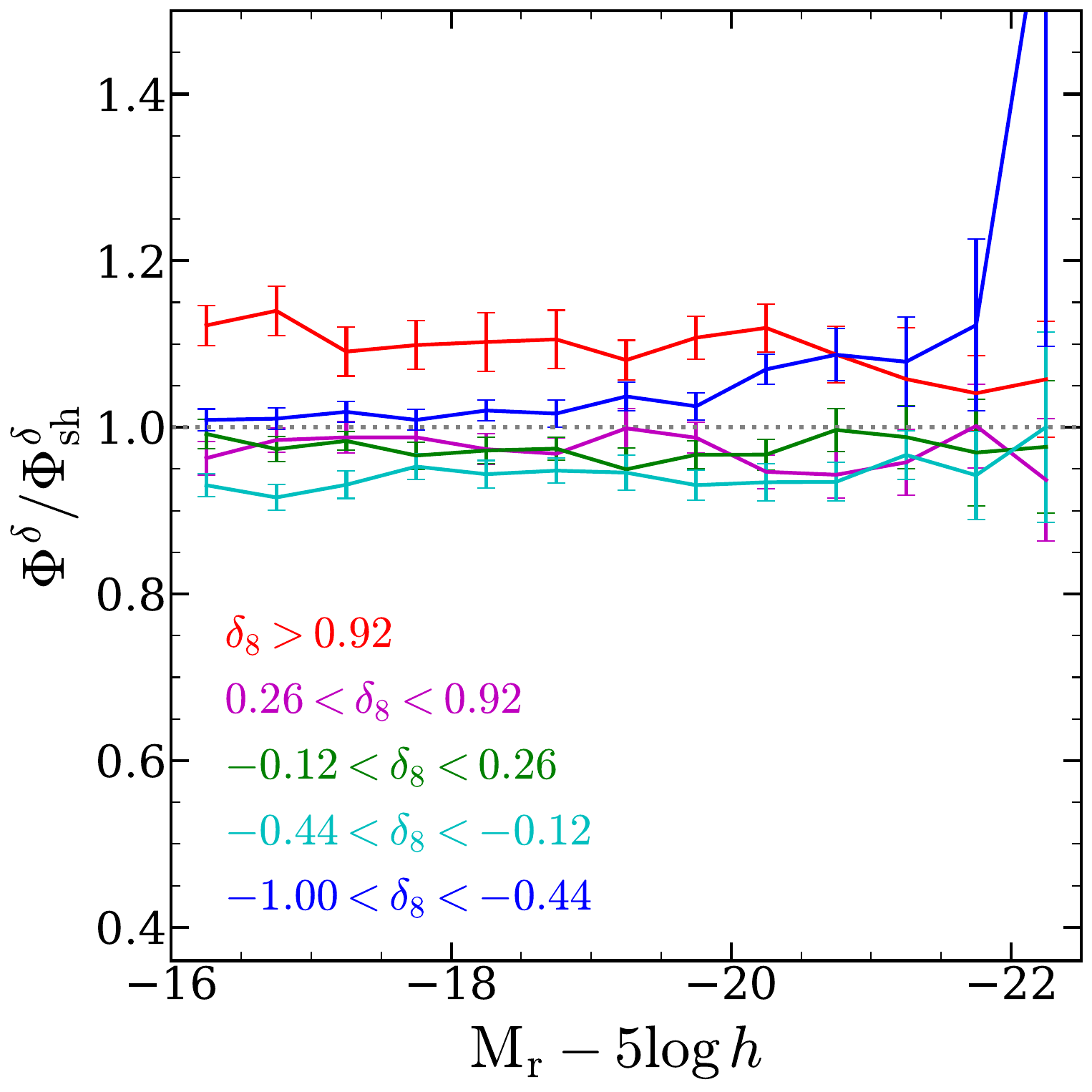}
\caption{The colored lines show the ratios of the LFs of our TNG300 full sample and the corresponding shuffled sample, for each density bin as labeled, when calculating the environment separately for each. Namely, we use $\delta_8$ measured for the original sample for the full sample LFs,  but recalculate $\delta_8$ from the shuffled sample and use it for the shuffled sample LFs. This is in contrast to our main results shown in the right panel of Fig.~\ref{fig:2}, where the original density measure was used consistently for both samples. The error bars are again derived from $27$ jackknife samples.}
    \label{fig:10}
\end{figure}

Figure~\ref{fig:10} shows the resulting measurement of the $\Phi^{\delta}/\Phi_{\mathrm{sh}}^{\delta}$ ratio, for the five density bins, when switching to the shuffled environment for the shuffled sample LFs. Assembly bias effects are still detected for the different density bins, but with less distinct trends. For the densest environment bin, we find a similar (slightly larger in fact) impact on the LF (namely a $\sim 10\%$ increase across all luminosities). However, in contrast to the right panel of Fig.~\ref{fig:2}, the density dependence of the signal is no longer monotonic, and in particular, there is much less signal for the least-dense bin. This was for one shuffled sample, but we find the same trends when repeating this measurement with several different shuffled samples (and their corresponding environments), implying that these deviations are not random.  We conclude that not using the same environment consistently can introduce systematics that change the measured signal.  

The physical explanation for these systematics can be traced to the impact of assembly bias. Environmental effects associated with assembly bias tend to further increase the number of galaxies in dense environments and decrease the number of galaxies in underdense regions, as discussed in \S~\ref{sec:result1}. Namely, it tends to make the different environments more `extreme', with larger absolute values of $\delta_8$. The shuffling procedure eliminates this effect, resulting in modified, smaller $\delta_8$ amplitudes in the extremes of the shuffled environment. Consequently, when using the same density thresholds as before now for the shuffled sample, less halos (and galaxies) will make the cut into the densest bin. This will amplify the assembly bias impact of more galaxies in the densest bin, relative to a shuffled sample. Similarly, less galaxies will make it into the least dense threshold for the shuffled sample, which in this case acts to diminish the measured assembly bias impact of less galaxies in underdense regions. The intermediate density bins will also be impacted by the changed $\delta_8$ values, resulting in less obvious trends. This provides a qualitative reasoning for the differences we see in Fig.~\ref{fig:10} compared to the right panel of Fig.~\ref{fig:2}. Stated in another way, the systematics arise from not having the identical set of halos in each density bin for the original sample and for the shuffled one, due to the specific changes in the measured density. In contrast, when using the same environment measure consistently for both original and shuffled samples (whether it be the dark matter density, the original galaxy density, or even the shuffled galaxy density), this systematic is naturally eliminated. 

\begin{figure}
\includegraphics[width=\textwidth]{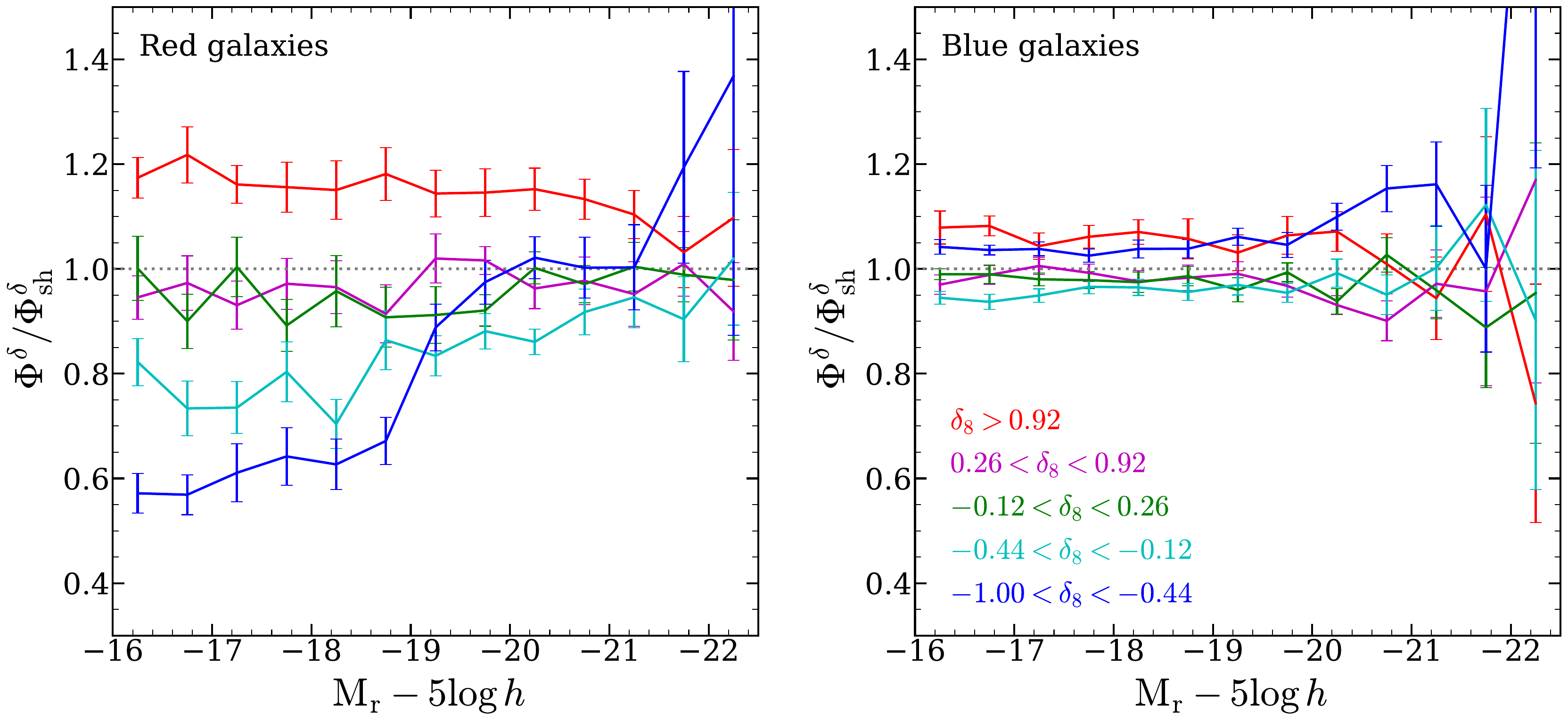}
\caption{Ratios of the LFs of original and shuffled samples in the different density bins, when calculating the environment separately for each, shown for the red galaxies (left panel) and blue galaxies (right panel). Specifically, we use $\delta_8$ measured from the TNG300 full sample for the original red and blue samples,  and recalculate $\delta_8$ from the shuffled full sample and use it for the shuffled LFs. This is to be compared to Fig.~\ref{fig:7}, where the original density measure was used consistently for both original and shuffled samples. The error bars in each panel represent jackknife errors.}
\label{fig:11}
\end{figure}

We also examine the impact of using the shuffled environment for the shuffled sample, when splitting the galaxy sample by color.  Figure~\ref{fig:11} shows the measured LF ratios in the different bins, separately for red and blue galaxies. Comparing this to Fig.~\ref{fig:7} (where the original environment is used consistently), we find systematics for the LFs of the blue galaxies (right panel) similar to those for the full sample.  For the red galaxies, however, we still get distinct signatures like before, as can be seen in the left panel of Fig.~\ref{fig:11}.  It appears that in that case the assembly bias signal is so strong so that, even with these systematics, we get a significant and robust measurement of it. In particular, the difference between the LF ratios of the densest bin to least dense bin (not shown) still exhibits about a factor 2 difference associated with assembly bias. These considerations are important to keep in mind when developing practical applications to model and detect assembly bias in observations of the environment-dependent LF, and when interpreting the results of such methods. Whenever possible it is advisable to use the same environment measure consistently, though when comparing models to real data one might invariably be facing such issues. Nonetheless, it is reassuring that, even with these concerns, our approach is able to produce a reliable signal for the faint red galaxies.

\section{The Halo Mass Function Dependence on Environment}
\label{sec:HMF}

\begin{figure}[b]
\centering 
\includegraphics[width=0.49\columnwidth]{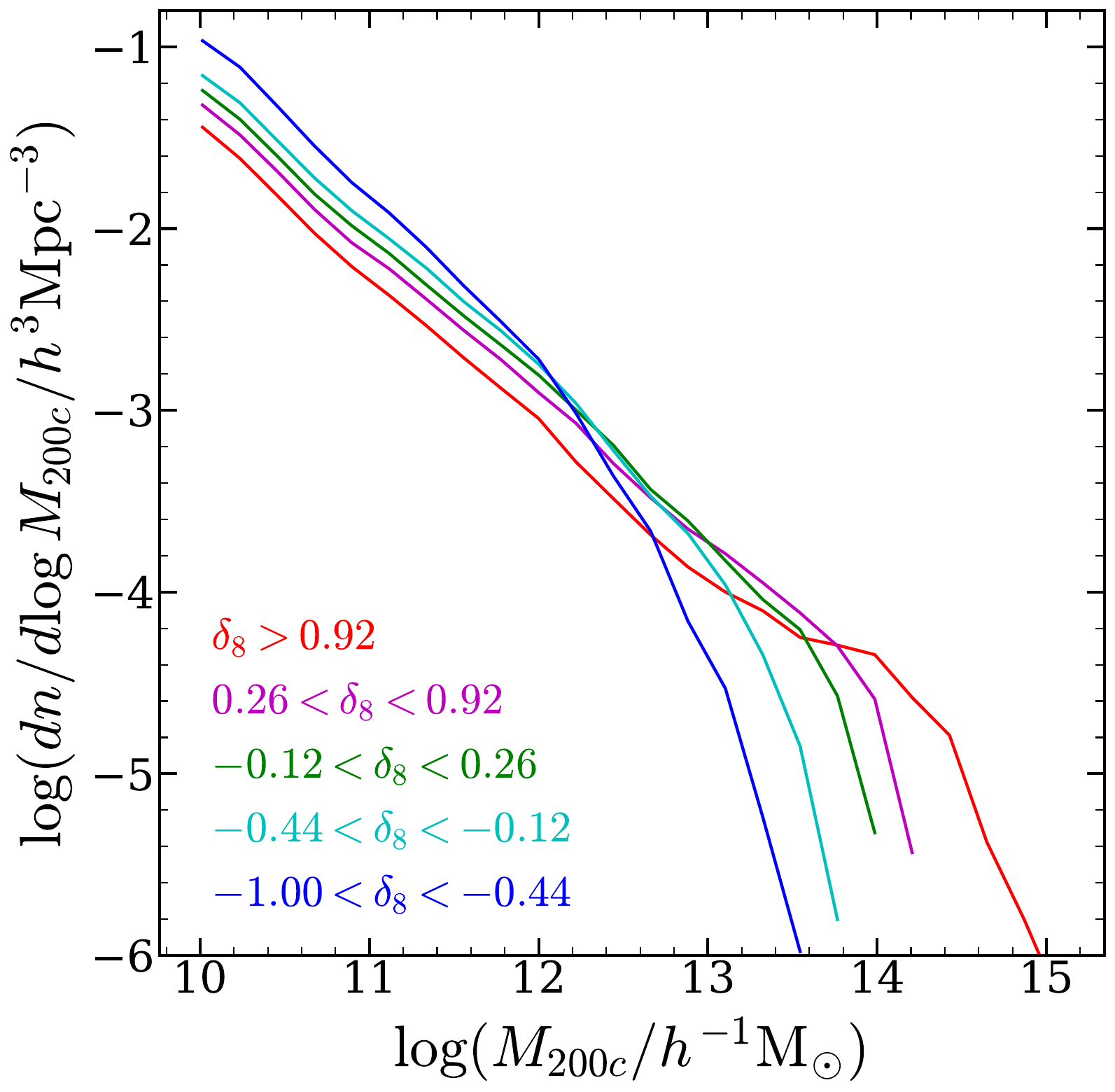}
\caption{The halo mass function dependence on environment for halos in the TNG300 simulation.  The colored solid lines represent the halo mass function measured for the halos in each of our five density bins, as labeled.}
\label{fig:12}
\end{figure}

The dependence of LF on environment is primarily driven by the halo mass dependence on environment.  While we focus in this paper on the secondary dependences due to assembly bias effects, it is insightful to consider also the halo mass dependence on environment.  Figure~\ref{fig:12} shows the halo mass function computed in the TNG300 at redshift $z=0$ for the five different environment bins used in our analysis. We use $M_{200c}$, defined as the mass within a radius enclosing a density contrast $\Delta = 200$ relative to the critical density $\rho_c$, as the virial mass of the halos. As shown, the halo mass function strongly depends on environment.  While its shape is largely independent of environment on the low-mass end, its magnitude varies. The amplitude of the halo mass function increases monotonically with decreasing density, such that there are distinctly more low-mass halos in low density regions. At the high-mass end, the halo mass function exhibits a sharp turnover. The mass scale of this turnover again depends strongly on environment, shifting to larger halo masses with increasing density. This leads to a significant prevalence of high-mass halos in dense regions and a paucity of them in less dense environments.

The dependence of the halo mass function on environment, coupled with the varying galaxy occupation with halo mass, leads to the fundamental dependence of the LF with environment. This is the origin of the variations of the LF with environment shown for the shuffled galaxy samples in the left-hand side of Fig.~\ref{fig:2}.  For example, the trend observed for brighter magnitudes -- beyond the break -- is consistent with the environmental trend seen at the high-mass end of the halo mass function in Fig.~\ref{fig:12}, and demonstrates that the environment dependence of the LF mainly arises from the dependence of halo mass on environment. Our results are in agreement with the findings of \citet{2004MNRAS.349..205M}, who model the environment dependence of the LF using a halo model without invoking assembly bias. See also related work by \citet{2015MNRAS.447.2683A} which examines how the halo mass function varies in different cosmic web environments.

\section{Additional Galaxy Distributions in a Slice of the Simulation}
\label{sec:slice}

\begin{figure*}[t]
  \includegraphics[width=0.95\textwidth]{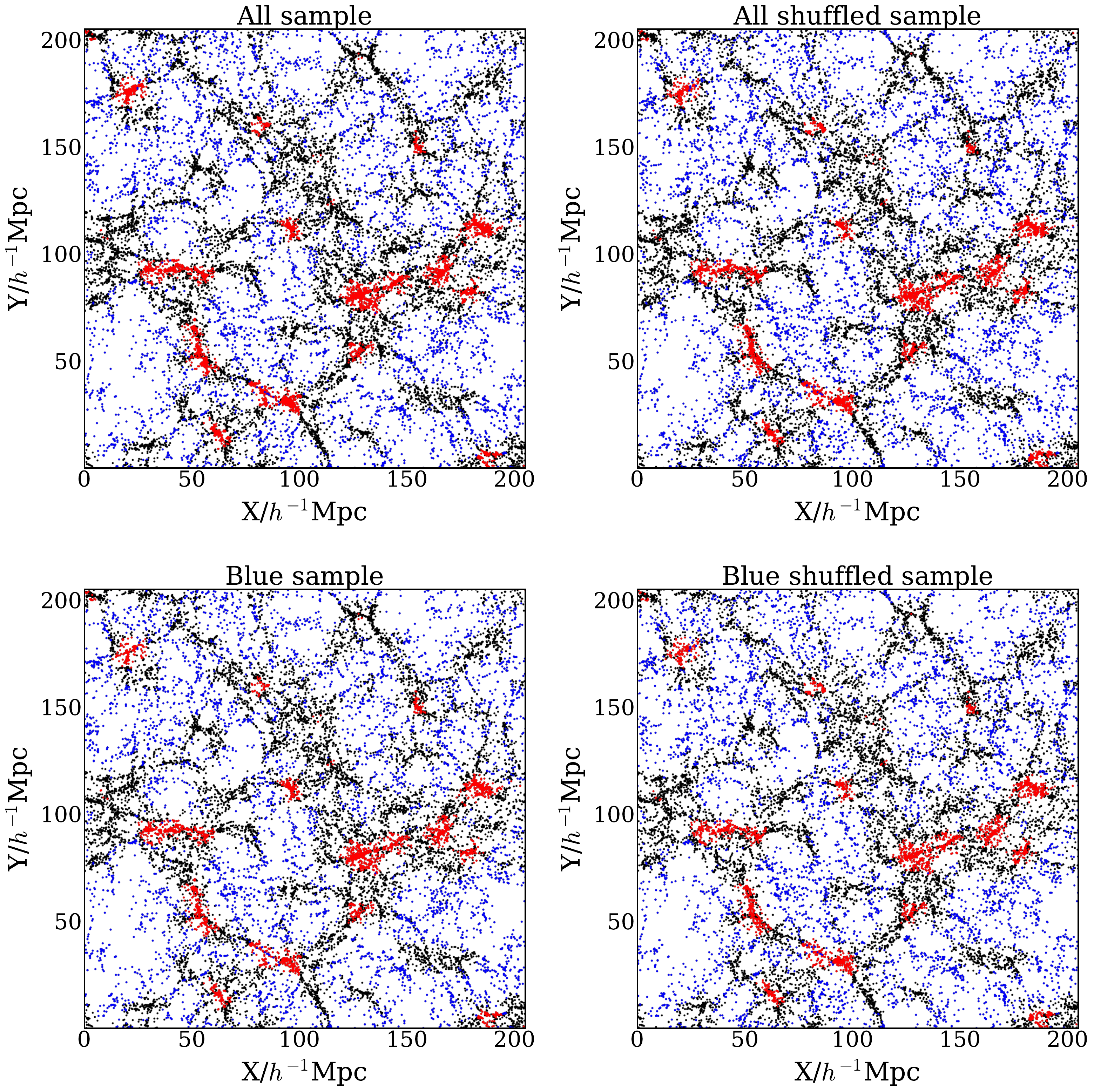}
\caption{The spatial distribution of faint galaxies in a $205 h^{-1} \text{Mpc} \times 205 h^{-1} \text{Mpc} \times 20 h^{-1} \text{Mpc}$ slice of the TNG300 simulation. The figures include all galaxies (top panels) and the blue galaxies (bottom panels) with $r$-band magnitude $-16.0> \mathrm{M_r} - 5 \log h> -18.5$. The left two panels show the original positions of the galaxies in the slice, while the right panels show the galaxies after (one realization of) shuffling the galaxy positions among halos of the same mass. The red dots represent the most dense regions ($\delta_8 > 0.92$), the blue dots represent the least dense regions ($-1.00 <\delta_8 < -0.44$), and the black dots show the remainder of the galaxies between these two extreme environments.} 
    \label{fig:13}
\end{figure*}

As discussed in \S~\ref{subsec:GAB_color}, Figure~\ref{fig:8} illustrates the impact of assembly bias on the environment-dependent LF in a slice from the simulation using the distribution of the faint red galaxies, which exhibit the strongest signal. In that case, the differences are large enough to be able to spot in visual inspection.  To complement that result, we present here the galaxy distribution in the same slice for all faint galaxies (red and blue) and for the blue ones separately. Figure~\ref{fig:13} shows the position of all galaxies (top panels) and just the blue galaxies(bottom panels) with $r-$ band magnitude $-16.0> \mathrm{M_r} - 5 \log h> -18.5$. The panels on the left show the original galaxy positions, while the ones on the right use their positions in one realization of the shuffling (the same realization as in Fig \ref{fig:8}). In all four panels, the red dots mark the galaxies in the densest environment bin with $\delta_8 > 0.92$, blue dots correspond to galaxies in the least dense bin with $-1.00 < \delta_8 < -0.44$, and black dots mark the remainder of the galaxies.

Comparing the top panels to the bottom ones, the distribution of all galaxies looks similar to that of the blue galaxies, which is as expected since the blue galaxies dominate the galaxy population at the faint end, as shown in Fig \ref{fig:5}. Comparing the left panels to the right ones, for both cases, the galaxies in the original sample appear overall a bit more clustered with more galaxies populating the dense environments (red dots) and less galaxies in the most underdense environments (blue dots). However, these assembly bias effects are subtle and significantly smaller than the ones exhibited for the faint red galaxies (as shown in the left panel of Fig.~\ref{fig:7} and in Fig.~\ref{fig:8}) and thus harder to discern visually.

\end{document}